# On the Lifshitz formula of dispersion interaction

M.V. Davidovich

Saratov National Research State University named after N.G. Chernyshevsky

E-mail: DavidovichMV@info.sgu.ru

The Lifshitz formula and methods of its preparation in the literature are considered. It is shown that in Lifshitz's work itself, this formula is given without a consistent conclusion. Moreover, the approach to the conclusion proposed in this work does not allow us to obtain it. The most general conclusion of this formula can be the method proposed by Levin and Rytov, the variation method of Schwinger and the method proposed by Van Kampen and co-authors. The Levin and Rytov approach is applicable in principle to bodies of arbitrary shape if the diffraction loss fields for electric and magnetic dipoles are determined, while the Van Kampen approach is applicable to any plane-layered structure and is quite simple. It is enough to write down the dispersion equations of the plasmon-polaritone structure. The specific dispersion force for a number of structures is calculated based on the Van Kampen method. It is shown that at small gaps, the force (pressure) density changes the inverse fourth-degree dependence on the distance and practically ceases to depend on it at distances less than 1 nm. For thin identical plates, this density is proportional to the square of their thickness at such distances, but the dependence quickly becomes saturated and already at thicknesses of the order of 10 nm practically ceases to depend on it.

## 1. Introduction

The Lifshitz formula for the dispersion force at zero temperature was first given more than 70 years ago in [1] without a derivation. Then this formula for an arbitrary temperature is given in [2], also without a derivation. In [3], an attempt was made to derive this formula using Maxwell's equations with fluctuation sources introduced in [4-7]. However, there is no consistent derivation of this formula in [3]. So, the formula (2.2L) from this work is given after the phrase "*By writing the squares of the integrals (1.9) in the usual way in the form of double integrals and performing one integration over the δ-functions, we obtain after some transformations ...*". Everywhere L means a reference to the formula from [3]. Next, we read: "*moreover, integral expressions from (1.13) should be substituted here as v, **w**, and the average products $g_i g_k$ should be understood simply as $(A\varepsilon''/4\pi 3)\delta_{ik}$.*" However, $\varepsilon''$ is not present anywhere in the final formulas, whereas according to [3] the force density should be proportional to $\varepsilon''$. Thus, the absence of dissipation should, according to the formulation in [3], mean the absence of force, which is incorrect. Next, we read "$dk_x$ integration is performed using the formula"

$$I(s) = i\pi / \left(|s|^2 (s^* - s)\right).$$

This formula is clearly incorrect. It is enough to indicate that its result does not give the correct dimension, which should be the inverse of $|s|$. The formula is proportional to $|s|^3$. Secondly, the result must be valid because

$$I(s) = 2\int_0^\infty \frac{dx}{\sqrt{(x^2 + s'^2)^2 + (2s's'')^2}}.$$

The correct formula according to the reference books [8,9] (formulas (3.165) and (1.2.75)) has the form $I(s) = F(\pi, k)/(2\pi)$, $k = \sqrt{\sqrt{(2s's'')^2 + s'^4} - s'^2 / \sqrt{(2s's'')^2 + s'^4}}$. With an imaginary value $s = is''$, it will be $I(s) = \pi/s''$, but not $I(s) = \pi/(2s''^3)$. The phrase "*after a series of transformations*" is found later in [3] and when giving the basic formula: "After a series of transformations, $F_\omega$ can be represented as follows": (2.3L). This final formula in [3] is correct. It can also be written as (2.4.L), as well as in the form [10]

$$P(d) = -\frac{\hbar}{2\pi^2} \int_0^\infty d\omega \coth\left(\frac{\hbar\omega}{2k_B T}\right) \text{Re} \int_0^\infty \kappa d\kappa k_{z2} \sum_{\alpha = e, h} \frac{r_1^\alpha r_3^\alpha \exp(-2ik_{z2}d)}{1 - r_1^\alpha r_3^\alpha \exp(-2ik_{z2}d)}. \qquad (1)$$

This is the Lifshitz formula [2,3] for the density of the attractive force $P(d) = F(d)/L^2$ between two half-spaces $j = 1$ and $j = 3$ with slit $j = 2$. It is written using the reflection coefficients $r_{1,3}^{e,h}$ of the modes from the half-spaces on the side of slit 2. This is a more general entry than in [3]. Here, as usual $k_{zj} = \sqrt{k_0^2 \varepsilon_j - \kappa^2}$, $\alpha = e, h$, and $r_j^\alpha = (y_2^\alpha - y_{in}^\alpha)/(y_2^\alpha + y_{in}^\alpha)$. $y_j^e = k_0 \varepsilon_j / k_{zj}$, $y_j^h = k_{zj}/k_0$ are the reflection coefficients and wave conductivities of the E-modes and H-modes from the gap side, and (1) corresponds to the internal Casimir pressure in it. Negative pressure means attraction. Obviously, the reflection coefficients cannot be proportional to e', the sums in (1) can be represented through a geometric progression (Neumann series) [10]

$$\left[1 - r_1^\alpha r_3^\alpha \exp(-2ik_{z2}d)\right]^{-1} = \sum_{n=1}^\infty r_1^\alpha r_3^\alpha \exp(2ik_{z2}d)^n.$$

Expression (1) is defined up to an infinite contribution, which is characteristically described by the following quote from [3]: "*Expression (2.2) is finite in itself, but contains terms that diverge when integrated over $d\omega$. This is the term c $\omega^3$, which occurs when integrating terms with 1/2 in curly brackets over dp. This divergent term, however, does not depend on the distance l between* *the bodies and therefore has no relation to the force of their mutual attraction that interests us and should be omitted. It represents the force of the reverse action of the bodies' own field on these bodies themselves, which is actually compensated by the same forces on other sides of the body.*" This is due to the fact that the force in [3] is defined in terms of the normal component of the Maxwell stress tensor from the slit side, but without taking into account the back pressure on the far surface (still considering the length of the dielectric plate to be finite). Note that in [3] a Gaussian system of units with time dependence $\exp(i\omega t)$ is used. We use the International System of Units. Accordingly, the introduction of a random field **K** as $-i\omega \mathbf{K}/c = 4\pi \mathbf{J}^0/c$ into (1.1L) requires performing correlation relations for the fluctuating external current $\mathbf{J}^0$ [3–7] $\langle J_l^0(\mathbf{r}), J_k^0(\mathbf{r}') \rangle = \omega^3 \varepsilon''(\omega) \delta_{lk} \delta(\mathbf{r}-\mathbf{r}') \Theta(\omega,T)/(4\pi^2)$. Here $\Theta(\omega,T)$ is the energy of the quantum oscillator [5]. In our case of SI system $\langle J_l^0(\mathbf{r}), J_k^0(\mathbf{r}') \rangle = \omega \varepsilon''(\omega) \delta_{lk} \delta(\mathbf{r}-\mathbf{r}') \Theta(\omega,T)/\pi$. According to [5], this dependence is obtained in the approximation for a very thick plate. Generally speaking, it should depend on the thickness and generally on the coordinates inside the body. Precisely, it is obtained from the principle of detailed equilibrium, internal fluctuation sources create in the far wave zone the same radiation density that is absorbed by the body due to the thermal field incident on it. It is very difficult to obtain such ratios. Relations in [4–7 are approximate for bodies. Lifshitz simplified the problem by considering the field only in the gap and taking the Maxwell stress tensor component (MST) $T_{xx}$ (formula (2.1L)) in it, which required the exclusion of infinite contributions. A strict formulation requires consideration of dielectric plates of finite thickness and crosslinking of fields at all four boundaries. In this case, standing modes arise in the plates, and outside the plates are the modes they emit (by radiated here we also mean modes with a surface character). It is important that one of the two types of dependencies $\exp(-ik_z z)$ and $\exp(ik_z z)$ is present (at $z>0$, the first one). In contrast to [3], we will take the *z* axis as the normal axis. Then in SI, the force density *F=P* in (1) should be understood as the reflection coefficients from the gap side of the plates in vacuum. Here we took into account the length of the structure t and took another term for $z=-t$. Further in [3], the function **K** for an infinite half-space is represented as a Fourier cosine integral (1.3L). It is not clear why the cosine is chosen for the asymmetric structure. Moreover, such a function does not decrease and has no limit at $x=-\infty$, i.e. the Fourier transform is invalid. It can be carried out for a finite thickness with a transition to the limit, but then the margins should also be stitched at opposite boundaries. At the same time, the correlation from (1.3L) clearly does not lead to a delta function with respect to the coordinates *x, x'* (these coordinates are negative, therefore, the modules should be taken under the cosines). If (1.4L) is true, then, for example,

$$\langle K_k(\mathbf{r}), K_k(\mathbf{r}')\rangle = 2A\varepsilon''(\omega)\delta_{lk}\delta(y-y')\delta(z-z')\int_0^\infty \cos(kx)\cos(kx')dk. \qquad (2)$$

The doubled integral has a limit $\sin(k(x-x'))/(x-x')+\sin(k(x+x'))/(x+x')$ for large $K$. At $|x+x'|\neq 0$ and $|x-x'|\neq 0$, it is not zero (as it should be) and does not even exist, and it does not define the delta function in the left half-space in any way. It goes to infinity at $|x|=|x'|$, i.e. both inside and outside the domain. As is known, from the momentum balance theorem [11, 12], the normal component of the MST gives (with a minus sign) a specific impulse flowing into the body, i.e. determines the negative pressure. Here we consider a body in a vacuum. To determine the pressure, it is convenient to consider the internal normal. In the case of a plate of finite thickness, Fig. 1, the total force density will be given by the pressure difference in the slit and outside the plate. Fig. 1 shows plates in a rectangular resonator with perfect walls. In this case, it is easy to write down the characteristic equations for complex frequencies and determine the sum for temperature zero by solving the equations for E-modes and H-modes:

$$\tilde{E}(d)=\frac{\hbar}{2}\mathrm{Re}\sum_{\substack{\alpha=e,h\\\beta=e,h}}\sum_{mnl}\tilde{\omega}_{mnl}^{\alpha\beta}(d). \qquad (3)$$

In the general case of dissipation the frequency is complex. Aiming as in [13] $L\to\infty$, let's move from summation by transverse indices to integration. In this case $dk_{xm}=(\pi/L)dm$, $dk_{yn}=(\pi/L)dn$. With an infinitely large size $L$, the indices run through continuous values, and the resonator modes turn into waves – plasmon-polaritones (PP). For integration, we introduce the polar coordinates $\kappa^2=k_{xm}^2+k_{yn}^2$, $dmdn=(L/\pi)^2 dk_{xm}dk_{yn}$, $dk_{xm}dk_{yn}=d\kappa d\varphi$. Angle integration gives $2\pi$. Summation by the longitudinal index can also be replaced by integration $dl=(D/2\pi)dk_{zl}$.

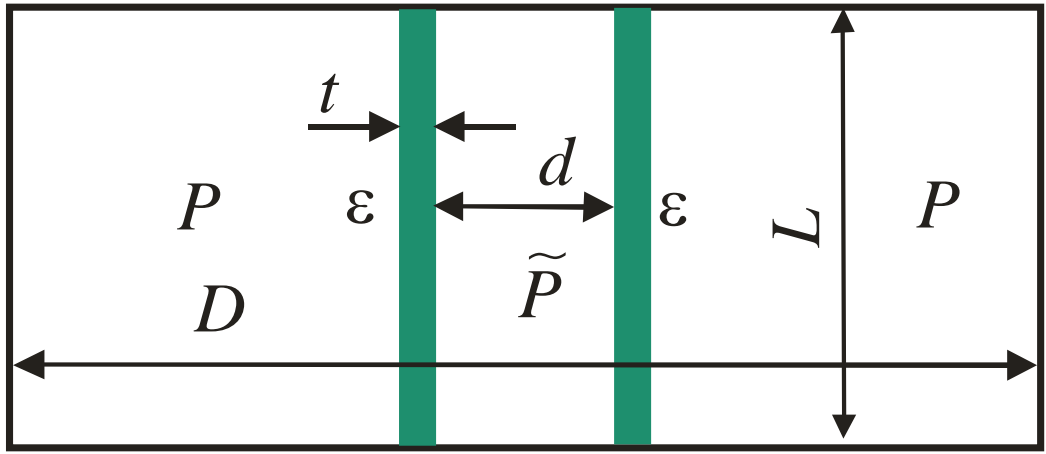

Fig. 1. Rectangular resonator with two dielectric layers with $L_x=L_y=L$ and $L_z=D$

At the finite temperature and the finite D (3) should be taken as

$$\tilde{E}(d) = \frac{\hbar}{2}\frac{L^2}{2\pi}\mathrm{Re}\int_0^\infty \sum_n \left(\tilde{\omega}_n^e \coth\left(\frac{\hbar\tilde{\omega}_n^e}{2k_B T}\right) + \tilde{\omega}_n^h \coth\left(\frac{\hbar\tilde{\omega}_n^h}{2k_B T}\right)\right)\kappa d\kappa. \quad (4)$$

The average energy of a quantum oscillator $\Theta$ is used here, and the frequencies are determined through discrete solutions $k_{zn}$ of the equations as $\omega = c\sqrt{\kappa^2 + k_{zn}^2}$. Assume that the losses in the dielectric in the all frequency range are small (infinitesimal), i.e. (5) includes the real parts of the frequencies. Next, we show that the result is also valid for dissipation. We have characteristic equations $f_{e,h}(\omega_n) = 0$ for frequencies. Passing to the limit $D \to \infty$, we obtain the dispersion equations $\omega^e = (\kappa, k_z)$ and $\omega^h = (\kappa, k_z)$. We will write them down as $f_{e,h}(\omega_n^{e,h}) = 0$. In infinite space, the frequency varies in a continuous spectrum.

According to [3], all field expansion coefficients must be proportional to $\varepsilon''\Theta$. We have a quote: "*The calculation leads to the following formulas for the components **v** and **w**, expressing them in terms of the amplitudes **g** of the «random" field»*". It follows that in the absence of dissipation ($\varepsilon'' = 0$) F=0, but this is not the case. The final formula is proportional $\Theta$ (like (1)), but not proportional $\varepsilon''$. This suggests that the correct formula (2.2L) and, accordingly, (1) could not be obtained within the framework of the approach outlined in [3] (even if the errors noted above were corrected). In particular, it follows the result of Casimir [13] for an absolutely non-dissipative structure, obtained by summing (3). It gives a finite force density for ideal dielectric plates $\varepsilon'' = 0$. Most likely, it was ingeniously guessed by the author [3], since it is quite reasonable to represent the field in the gap as the sum of multiple delayed reflections [10].

Thus, both formula (1) and the corresponding formula (2.2L) cannot be obtained as solutions to boundary problems with fluctuation sources inside bodies satisfying relations (2). In [5], these formulas were obtained by finding the mixed diffraction loss of field when the structure is excited by single external electric and magnetic dipoles located outside the bodies. It is noteworthy that the field correlations in formulas [5] are determined through the mixed diffraction losses of single sources (electric and magnetic dipoles) located outside the bodies (formulas (5.5), (5,6), (15.7) – (15.9) from [5]). In this case, it is possible to obtain dispersion forces at different body temperatures and in the nonequilibrium case, when the radiation temperature is different from the body temperatures. Generally speaking, internal correlation sources are not known for the formulation of the problem according to [3]. They need to be found by solving the problem of diffraction of the thermal field and radiation of a body in the far zone (see, for example, [14]). In general, they should depend on the shape of the body and its coordinates, i.e. on the location of the fluctuations. The sum of the radiation reflected by the body and the radiation from these sources should give the total equilibrium density of thermal radiation. These are complex inverse problems. The use of the ratios [4–7] is approximate, especially for thin plates. In [15–18], the conclusion was made on the basis of the variational principle, and an external polarization source, which defines the Green's function (GF), and the variation of the dielectric permittivity (DP) were also considered. Formula [3] was obtained in [16] in two ways: directly by minimizing of energy and based on the Maxwell stress tensor (MST) with the determination of field correlations in terms of GF.

So, in the plane-layered structures under consideration, PP occurs along the boundaries of the dielectric. They are divided into fast leakage (flowing) and slow surface ones. The dispersion of DP leads to the fact that they are complex. In this case, fast modes are radiated, while slow ones exponentially decay in the normal direction. In the case of ideal screens in the $z$ direction, the resulting modes form spatial resonances with complex frequencies described by characteristic equations. By removing the screens, we obtain open structures, and the characteristic equations turn into a dispersion equation (DE) for PP with continuously varying frequencies. Slow PP make the main contribution to the dispersion force at short distances, and at large distances they decay exponentially, whereas fast radiated modes are significant at large distances. The role of PP in the creation of dispersion forces is considered in [19, 20] and in the literature cited there. The formal absence of losses leads to actual DE having, generally speaking, bands in which PP is absent. Taking PP into account in the Lifshitz structure requires the use of the stitching (mode matching) method at least at four boundaries.

The absence of consistent derivation of the formula in [3] led to the fact that in many works formula (1) was repeatedly derived. Here we should mention the works [21–23], which used the methods of quantum field theory and quantum statistical physics with GF approach, as well as the monograph [5]. In it, the approach proposed in [3] "*using a direct technique – by solving an inhomogeneous boundary value problem with distributed random sources in 1 and 2 media*" according [5] was not used. As will be seen later, it cannot be used. Instead, [5] developed a universal approach based on obtaining the fields of point dipoles outside the bodies (in the gap) and using the Lorentz lemma to determine the correlations. Next, the diffraction fields in the gap and the force density are determined through the correlations of the Maxwell tensor. Namely, let $\mathbf{J}_1 = \mathbf{l}_{1(e,h)}\delta(\mathbf{r}-\mathbf{r}_0)$ and $\mathbf{J}_2 = \mathbf{l}_{1(e,h)}\delta(\mathbf{r}-\mathbf{r}_0)$ be the unit current densities of point electric and/or magnetic dipoles in the gap. For this purpose, special heat for both dipole's fields losses are used (with including of diffraction). Lorentz's lemma leads to correlations $\langle E_{\mathbf{l}_1}(\mathbf{r}_1), E_{\mathbf{l}_2}(\mathbf{r}_1)\rangle = -\Theta(\omega,T)\mathrm{Re}(\mathbf{l}_2 \mathbf{E}_{01})/\pi$, $\langle H_{\mathbf{l}_1}(\mathbf{r}_1), H_{\mathbf{l}_2}(\mathbf{r}_1)\rangle = -\Theta(\omega,T)\mathrm{Re}(\mathbf{l}_2 \mathbf{H}_{01})/\pi$, where zero indicates the fields of the corresponding dipoles that create mixed diffraction losses (formulas (15.7), (15/8), p. 167 in [5]). The fields of dipoles are actually nothing more than GFs. The method is quite universal. For the Lifshitz's problem, in addition to the contribution (2.3L), he allocates the pressure of the thermal field on the plate $P_\omega = \Theta(\omega,T)\omega^2/(3\pi^2 c^3)$ (for vacuum). Here $u = 4\varpi_{SB}\Theta(\omega,T)\omega^2/c$ is the density of the thermal radiation and P=u/3 is it's the pressure. In thermal equilibrium, at each frequency, the body absorbs as well as radiates, so this pressure is the same as on an ideal mirror [5]. The same pressure occurs on the other side of the plate. However, the method allows us to consider different body temperatures and thermal fields.

The proof of formula (1) was also obtained in the works of Schwinger [15–18] on the basis of methods of field source theory. The strength is determined based on the variation of the electromagnetic field caused by external polarization sources. The fields are found based on the GFs method. In [16], the inaccuracy of the Casimir force correction for real metal plates allowed in [3] is considered. A similar inaccuracy of the temperature correction (5.4L) is considered in [5], where a formula with a different sign and coefficient is obtained. The Lifshitz formula and a similar formula from [23] were also obtained in the work of Van Kampen et al. [24]. Its generalized conclusion in case of delay is given in the work of Schram [25]. A generalization to the case of finite temperature is considered in [26]. Of all the methods considered, this method seems to be the most universal and simplest. The final formula is immediately written in a standard form using the DE for PP structures, which are easy to find. Also, the correlation relations for fields written in terms of GFs are given, for example, in [27], [28]. There are other publications on the derivation of formula (1), mainly derived from the definition of free energy. Note that the form (1) does not restrict the configuration as a gap between half-spaces. Plane-layered structures can be considered both to the left of the slit and to the right, and formula (1) determines the force density between them. Next, we will consider in more detail the Van Kampen method [24,25]. It implies the summation of the poles corresponding to the zeros of $f_e$=0 and $f_h$=0 based on the principle of the argument. It is also considered in [29, 30]. The DE for PP can be written as a function of the frequency of the wave vector. For continuous indexes in an open structure, surfaces in **k**-space are determined. The zeros of DU give the relationship between frequency and the wave vector **k**, i.e. modes along the surfaces (x,y). The effect of these modes on the Casimir force is considered in [19, 20], where summation of type (3), (4) for dissipative systems is actually considered. Note that to obtain the density of the attractive force, the energy should be differentiated: $P = -\partial_d E(d)$. The negative derivative means attraction.

## 2. Generalization of the Lifshitz formula by the Van Kampen method

The Van Kampen formula takes into account the sum of the energies $\hbar\omega_n/2$ as an integral over frequency and gives the Casimir pressure in the gap as [24,25,29,30]

$$P(d) = -\frac{\hbar c}{2\pi^2} \int_0^\infty \kappa d\kappa \int_0^\infty K_0 \left( \frac{1}{f_e(\kappa,k,d)} + \frac{1}{f_h(\kappa,k,d)} \right) dk. \qquad (5)$$

The zeros $\omega_n$ (discrete and continuous) correspond to the poles of the meromorphic functions in (5). Equality (5) at zero temperature is obtained from the principle of the argument in the form

$$\left[ \sum_n \omega_n(\kappa) \right]_{f_{e,h}(\omega_n)=0} = \frac{1}{2\pi i} \oint_C \frac{f'_{e,h}(\omega)}{f_{e,h}(\omega)} \omega d\omega = \frac{-1}{2\pi} \oint_C \left[ \ln(f_e(\omega)) + \ln(f_h(\omega)) \right] d(i\omega) =$$
$$= \frac{c}{2\pi} \int_{-\infty}^\infty \left[ \ln(f_e(\kappa,k,d)) + \ln(f_h(\kappa,k,d)) \right] dk = \frac{c}{\pi} \int_0^\infty \left[ \ln(f_e(\kappa,k,d)) + \ln(f_h(\kappa,k,d)) \right] dk \qquad (6)$$

Here $f_{e,h}(\omega_n) = 0$ are the characteristic equations or DE describing the resonant frequencies of the structures. These frequencies for dissipative structures lie in the upper half-plane of the complex plane $\omega$ symmetrically relative to the imaginary axis, i.e. $\omega_n = \omega'_n + i\omega''_n$, $\tilde{\omega}_n = -\omega'_n + i\omega''_n$, which corresponds to absorption [31]. Note that any actual damped oscillation can be represented using frequencies $\omega_n = \omega'_n + i\omega''_n$ and $\tilde{\omega}_n = -\omega'_n + i\omega''_n$. In the lower half-plane, the frequencies have negative imaginary parts, which corresponds to radiation. In thermodynamic equilibrium, the atoms of matter at each frequency absorb and emit equally. There is no energy accumulation or change in the system at all. In addition, the frequencies determine the energy in the gap and outside the plates, i.e. in a vacuum. The energy density in a vacuum is known and can be represented as a set of oscillator energies with frequencies $\omega_n = \omega'_n + i\omega''_n$. Next, it will be shown that the presence of a dielectric in the gap leads to the same formulas. We draw the contour so that (3) gives twice the sum of the positive frequencies, while the imaginary parts disappear, and the real part "Re" does not need to be taken. To do this, the contour should be drawn along the imaginary axis and closed around the circle in the right half-plane. At zero temperature we multiply (3) by $\hbar/2$. In general, it should be multiplied by $\Theta(\omega,T) = \hbar \coth(\hbar\omega_n/2k_B T)/2$. Then poles appear on the real frequency axis (or on the imaginary axis in the plane $\xi = i\omega$). In the plane $\xi$, we can take the contours as shown in Fig. 3.7 of [10] and obtain the formula (5.2L) or (21.48) from [10]. Taking the energy density in the form

$$\tilde{E}(d) = \frac{\hbar}{2} \frac{L^2}{2\pi} \mathrm{Re} \int_0^\infty \sum_n \left( \tilde{\omega}_n^e \coth\left( \frac{\hbar\tilde{\omega}_n^e}{2k_B T} \right) + \tilde{\omega}_n^h \coth\left( \frac{\hbar\tilde{\omega}_n^h}{2k_B T} \right) \right) \kappa d\kappa, \qquad (7)$$

moving from frequency summation and integration at $D \to \infty$, using $P(d) = -\partial_d \tilde{E}(d)/L^2$, we obtain (5). In open space, the frequency spectrum becomes continuous, and the characteristic equations turn into DE. Thus, the Van Kampen formula gives exactly the Lifshitz result. This formula at zero temperature has the form [29] (5), as it is usually used, and in the general case, a multiplier $\cot(\hbar ck/(2k_B T))$ should be added:

$$P(d) = -\frac{\hbar c}{2\pi^2} \int_0^\infty \kappa d\kappa \int_0^\infty K_0 \cot(\hbar ck/(2k_B T)) \left( \frac{1}{f_e(\kappa,k,d)} + \frac{1}{f_h(\kappa,k,d)} \right) dk.$$

According to [29], this formula is also determined up to a certain contribution independent of $d$ and corresponding to the infinite vacuum energy. For the Casimir problem $f_{e,h}(\kappa,k,d) = \exp(2K_0 d) - 1$, and $P(d) = -\hbar c\pi^2/[240 d^4]$ follows from (5). The result (5) is always real, whereas formula (2.4L) was originally based on taking the real part. An interesting quote [3] here is "*It is essential that it is possible to represent $F_\omega$ as the real part of the integral of the analytical function p, despite the fact that expression (2.2) is obtained by taking the squares of the modules of the field components*". The following inaccuracy in [3] is as follows: "*The poles of the integrand expression could be the denominators in (2.4), i.e. the roots of the equations*," and the following is the formula (2.5L) for DE, in which lacks $-1$. It is stated that the integral expression in (2.4L) has no poles in the complex plane. However, this is not the case: taking into account the lost of $-1$, these poles arise, and (2.4L) and (5) give the amounts of contributions from these "poles". With $\varepsilon'' = 0$ the DE is real, and the actual force density is determined only by it and is finite. In this sense, the Lifshitz formula actually means summation (3) with a continuous spectrum of real frequencies. When switching to the imaginary frequency, DP is real and is determined integrally through $\varepsilon''(i\omega)$. The quote is characteristic here: "*Thus, we can say that the law of interaction of bodies is completely determined by the assignment of their functions $\varepsilon''(i\omega)$ (we will see in sec. 5, that this remains true even at temperatures other than zero)*". It is determined integrally through reflection coefficients, whereas its proportionality to $\varepsilon''$ was assumed in the initial formulation. There really are no media with $\varepsilon'' = 0$. In the case of ideal media (for example, at zero temperature), the $\varepsilon''$ in Lorentz model is proportional to sets of delta functions with resonant frequencies $\omega_n$. This corresponds to the endless lifetimes of the levels. For such an ideal dielectric, formula (2.10L) gives

$$\varepsilon(i\xi) = 1 + \sum_n \omega_n^2/(\omega_n^2 + \xi^2).$$

At low frequencies (which determine the force at large distances), the spectral DP is real and independent of frequency (in the absence of free charges). The transmission of an electromagnetic pulse to the body is possible in two ways: due to reflection (the reflection coefficient plays a role here) and due to the absorption of photons entering the body [11,12]. In this example, reflection plays a major role. Formula (1) is, in fact, the result of summing the energies of oscillators (3), (4), performed at the limiting transition from the resonator to the open space, as a result of which frequency integration occurs.

There has been a long-standing dispute about whether relations (3) and (4) can be used to determine the Casimir energy in dissipative structures [19,20,29,32]. Within the framework of the open systems approach, the Lifshitz formula was derived in a number of papers [33–38]. In fact, the Lifshitz formula corresponds to the integral sum of modes of a dissipative system. There is a reason why Lifshitz initially took the real part in his formula. The Van Kampen formula actually calculates this real part by switching to the imaginary frequency axis when the imaginary parts cancel each other out. This is due to the implementation of the principle of detailed equilibrium for each frequency. It is characteristic that the Lifshitz formula is strictly proven for a non-dispersing medium in the gap [5]. However, the Van Kampen method also gives results when filling the gap with a dielectric with dispersion. Taking the real part in expressions like (3) can be explained by the fact that there is a stationary equilibrium energy density of the resonator is $\tilde{E}(\omega) = \varepsilon_0 \varepsilon'(\omega) |\mathbf{E}(\omega)|^2 / 2$, if there is no accumulated kinetic energy of charges moving under the action of the field [39–43]. In the presence of oscillating charges, for example, in plasma), their kinetic energy must be taken into account [39], and then the density of the stored average energy over the period (including kinetic energy) has the form [44]

$$\tilde{E}(\omega) = \frac{\varepsilon_0 |\mathbf{E}|^2}{4} \left( 1 + \frac{\omega_p^2}{\omega^2 + \omega_c^2} + \sqrt{\left(1 - \frac{\omega_p^2}{\omega^2 + \omega_c^2}\right)^2 + \frac{\omega_p^4 \omega_c^2}{(\omega^2 + \omega_c^2)^2 \omega^2}} \right),$$

which can also be represented as a set of oscillator energies (1), and at a low collision frequency (CF) $\omega_c$ we get $\tilde{E}(\omega) = \varepsilon_0 |\mathbf{E}(\omega)|^2 / 2$. An anachronistic formula holds for Lorentz oscillators [39]. Although it is possible $\varepsilon'(\omega) < 0$, after quantization, the field looks like a set of oscillators.

**3. Results for some structures**

Let us first consider the application of formula (1) for graphene sheets on wafers. In the absence of graphene, the input conductivity of such very thick a plate is $y_{in}^\alpha = y_j^\alpha$. In the presence of graphene $y_{in}^\alpha = y_j^\alpha + \varsigma$, where $\varsigma$ is the normalized conductivity of graphene $\varsigma = (\mu_0 / \varepsilon_0)^{1/2} \sigma$. We believe that the graphene sheets are fixed on the plate. For a symmetric structure we have $r_1^{e,h} = r_3^{e,h} = (y_0^{e,h} - y_{in}^{e,h})/(y_0^{e,h} + y_{in}^{e,h})$, where zero corresponds to a vacuum. Taking into account the thickness $t$ is not a problem, at the same time there are members with $\tan(tk_z)$. With a vacuum gap between free graphene sheets, we obtain the coefficients $r^e_{1,3} = -\varsigma/(2k_0/k_z + \varsigma)$, $r^h_{1,3} = -\varsigma/(2k_z/k_0 + \varsigma)$, $k_z = \sqrt{k_0^2 - \kappa^2}$. In this case, the sum in (1) is

$$\frac{\varsigma^2}{(2y^e + \varsigma)^2 \exp(-2ik_z d) - \varsigma^2} + \frac{\varsigma^2}{(2y^h + \varsigma)^2 \exp(-ik_z d) - \varsigma^2}.$$

When moving to the imaginary frequency $\omega = -i\xi$, we have $k_{zj} = -iK_j$, $K_j = \sqrt{k^2 \varepsilon_j + \kappa^2}$, $k = \xi/c$, $y^e = 1/\rho^e = k/K$, $y^h = 1/\rho^h = K/k$, $K_0 = K_2 = \sqrt{k^2 + \kappa^2}$, and

$$P(d) = -\frac{\hbar}{2\pi^2} \int_0^\infty d\xi \cot\left(\frac{\hbar\xi}{2k_B T}\right) \operatorname{Re} \int_0^\infty \kappa d\kappa K_2 \sum_{\alpha=e,h} \frac{r_1^\alpha r_3^\alpha \exp(-2K_0 d)}{1 - r_1^\alpha r_3^\alpha \exp(-2K_0 d)}. \tag{8}$$

The decomposition of the fraction in (8) into a geometric progression (Neumann series) represents the interaction as endless acts of re-reflections. For zero temperature, there is no cotangent in formula (8). At a finite temperature, this ratio can be expressed in terms of the sum of the Matsubara frequencies [2,3,10]. It should be noted that formula (8) exactly coincides with the Van Kampen summation formula [29], if $\hbar\omega_n/2$ is replaced by the average energy of the oscillator $\theta(\omega_n) = \hbar\omega_n \coth(\hbar\omega_n/2k_B T)/2$. Assuming that the collision frequency is zero at zero temperature, we obtain a purely imaginary $\varsigma$, whereas the correlations of the surface current density are proportional to $\varsigma'=0$. At the same time $r^e_{1,3} \neq 0$. For the Lifshitz problem with a vacuum gap between identical half-spaces $K = \sqrt{k^2\varepsilon + \kappa^2}$, and

$$f_h(\kappa, k_0) = \exp(2K_0 d)\left(\frac{K + K_0}{K - K_0}\right)^2 - 1, \tag{9}$$

$$f_e(\kappa, k_0) = \exp(2K_0 d)\left(\frac{K + \varepsilon K_0}{K - \varepsilon K_0}\right)^2 - 1. \tag{10}$$

The dispersion equations can be obtained for any plane-layered structure with an arbitrary number of layers and conductive sheets, for which it is simply necessary to use the stitching method or a similar method for the corresponding electrodynamic problem of determining eigenfrequencies. In this case, the force between any layers can be determined. In particular, [24] provides a formula corresponding to [23] for a dielectric layer between different half-spaces, obtained using the method of GFs. For two dielectric layers of thickness $t$ with a distance $d$ between them, we have [45]

$$f_h(\kappa, k_0, d, t) = \exp(2Kd)\left[\frac{2K\tilde{K}\coth(\tilde{K}t) + \tilde{K}^2 + K^2}{\tilde{K}^2 - K^2}\right]^2 - 1, \tag{11}$$

$$f_r(\kappa,k_0,d,t) = \exp(2Kd)\left[\frac{2\varepsilon K\tilde{K}\coth(Kt) + \tilde{K}^2 + \varepsilon^2 K^2}{\tilde{K}^2 + \varepsilon^2 K^2}\right]^2 - 1, \qquad (12)$$

They are obtained by stitching. Here $K_0 = \sqrt{\kappa^2 - k_0^2}$, $K = \sqrt{\kappa^2 - k_0^2\varepsilon(\omega)}$. It is convenient to designate $k = ik_0$ and switch to the imaginary frequency $\xi = i\omega = ck$. Then $K_0 = \sqrt{\kappa^2 + k^2}$, $K = \sqrt{\kappa^2 + k^2\varepsilon(-ick)}$, $\tilde{K} = \sqrt{\kappa^2 + \varepsilon k^2}$. The Lifshitz case corresponds to a thick layer with $\coth(\tilde{K}t) = 1$. For small plate thicknesses, the force density is proportional to the product of the thicknesses $t_1 t_2$. However, these thicknesses are already comparable to the original dimensions. Further, to simplify the formulas, we consider the plates to be the same. For the force density we should take $\partial_d \tilde{E}(d)/(L_x L_y)$, where $L_x L_y$ is the large area of the layers. The dielectric material is considered to be the same and has a spectral DP $\varepsilon(\omega) = \varepsilon'(\omega) - i\varepsilon''(\omega)$, moreover $\varepsilon''(0) = 0$, if there are no free charge carriers (for plasma $\varepsilon''(0) = \infty$). The tilde indicates the frequencies, energy, and longitudinal wavenumber perturbed by the dielectric. In an empty resonator with dimensions $L_x$, $L_y$, $L_z$, there are undisturbed resonant frequencies $\omega_{mnl}^{e,h} = c\sqrt{k_{xm}^2 + k_{yn}^2 + k_{zl}^2}$ of TE$_{mnl}$ (or H$_{mnl}$) modes (for index $e$), where $k_{xm} = m\pi/L_x$, $k_{yn} = n\pi/L_y$, $k_{zl} = l\pi/L_z$, $m=0,1,...$, $n=0,1,...$, $l=1,2,...$, except for $m=n=0$, as well as frequencies of TM$_{mnl}$ (or E$_{mnl}$) modes (for index $h$), the difference is which is that now $m=1,2,...$, $n=1,2,...$, $l=0,1,2,...$ [40,41]. Thus, oscillation degeneracy takes place in an empty resonator. In a filled resonator, it is removed: $\tilde{k}_{zl} = k_{zl} + \Delta k_{zl}$ is the value perturbed by the dielectric, $\Delta k_{zl} \sim 1/L_z$. Going to the limit $L_{x,y} \to \infty$ means continuity of the transverse indices $dk_{xm} = dk_x = (\pi/L_x)dm$, $dk_{yn} = dk_y = (\pi/L_y)dn$, and replacing the two-dimensional sum in (1) with a two-dimensional integral. It is convenient to switch to the polar coordinates $k_x = \kappa\cos(\varphi)$, $k_y = \kappa\sin(\varphi)$. Then the angle integral is calculated and is equal to $2\pi$. The transition to the limit $L_z \to \infty$ reduces the sum of the expansion of the resonator to the entire space to a two-dimensional integral over $dk_z d\kappa$, $dk_z = dk_{zl} = (\pi/L_z)dl$. The frequencies in the finite resonator are discrete. As noted, they lie in the upper half-plane of the complex frequency plane symmetrically relative to the imaginary axis $\pm\omega_n' + i\omega_n''$ [31]. The radiation corresponds to a change in the sign of the imaginary part, i.e. an oscillation increasing over time $\cos(\omega_n't)\exp(\omega_n''t)$. These frequencies lie in the lower half-plane. In a resonator with infinite walls and with a finite size $L_z$, the characteristic equations define frequencies $\tilde{\omega}_l = \tilde{\omega}_l(\kappa)$ as continuous meromorphic functions $\tilde{\omega}_l = \tilde{\omega}_l(\kappa)$, l=1,2,.... For plates in free space, the characteristic equations $f_{e,h}(\kappa, \tilde{k}_z) = 0$, which are functions of two variables, are the DEs of PPs. In a vacuum $\kappa^2 = k_0^2 - k_z^2$, but in a dielectric $\kappa^2 = k_0^2 \varepsilon - k_z^2$. The value $\tilde{k}_z$ in the structure is determined from the DE. It can be (considering the dissipation to be extremely small) real $\tilde{k}_z < k_0$ (fast leakage PPs, or modes radiated in vacuum), and imaginary, which determines slow PP along the surface. The frequencies perturbed by the dielectric are defined as $\tilde{\omega} = c\sqrt{\kappa^2 + \tilde{k}_z^2} = ck_0\sqrt{1 + (2k_z + \Delta k_z^2)/k_0^2}$. Then we can consider DE as a function of $\kappa$ and $k_0$: $f_{e,h}(\kappa, k_0)$, $k_0 = \omega/c$. Note that there are several possible forms of the characteristic equation, which are also the essence of DE for PPs [46]. The force density or Casimir pressure is defined as $P(d) = -\partial_d \tilde{E}(d)$. According to the method (see also [29,45]), we have equation (5) (in [29,45]), the multiplier 2 was lost, although then it was restored). The contour in the plane $\xi$ can be drawn as in Fig. 3.7 of Ref. [10], while the real parts of the frequencies are taken into account twice. When the average energy $\Theta$ is calculated, additional poles appear on the imaginary axis $\xi$, as a result of which the force density is calculated using the sum of the Matsubara frequencies. However, the sum is determined up to the multipliers $A_{e,h}(\kappa, k)$, because for any multiplier $A_{e,h}^{-1}(\kappa, k) f_{e,h}(\kappa, k, d) = 0$. The multipliers should be determined from the condition that in the limiting case $\varepsilon \to \infty$ the Casimir problem is obtained. So, for (2), (3) we obtain the Casimir problem $f_e(\kappa, k, d) = f_h(\kappa, k, d) = \exp(2K_0 d) - 1$, $P(d) = -\hbar c \pi^2/(240 d^4)$, i.e. in this case $A_{e,h}(\kappa, k) = 1$. In the absence of plates ($t=0$ or $\varepsilon = 1$) we have $f_{e,h}(\kappa, k, d) \to \infty$ and $P(d) = 0$. With a small plate thickness, we have

$$f_h^{-1}(\kappa, k_0) \approx t^2 (K^2 - K_0^2)^2 / [4K_0^2 \exp(2K_0 d)],$$
$$f_e^{-1}(\kappa, k_0) = t^2 (K^2 - \varepsilon^2 K_0^2) / [4\varepsilon^2 K_0^2 \exp(2K_0 d)],$$

and the force is proportional to the square of the thickness. Considering plates of different thicknesses leads to a proportional $t_1 t_2$ force. For the Casimir problem $\varepsilon \to \infty$, $K \to \infty$, and $f_e(\kappa, k_0) = f_h(\kappa, k_0) = \exp(2K_0 d) - 1$.

The DP model should be used to obtain numerical results. The dispersion of real dielectrics over a wide range is usually quite complex. Using Lorentz's law of dispersion, taking into account the internal field, it can be represented as the Clausius-Mossotti formula

$$\varepsilon(\omega) = \frac{1 + \frac{2}{3} \sum_{n,m=1} \frac{\omega_{pm}^2}{\omega_{mn}^2 - \omega^2 + i\omega_{cmn}\omega}}{1 - \frac{1}{3} \sum_{n,m=1} \frac{\omega_{pm}^2}{\omega_{mn}^2 - \omega^2 + i\omega_{cmn}\omega}}. \tag{13}$$

Here we have used Lorentz polarizabilities for an atom with transition frequencies $\omega_{mn}$

$$\alpha_{mn} = \frac{e^2 N_m}{\varepsilon_0 m_e} \frac{1}{\omega_{mn}^2 - \omega^2 + i\omega\omega_{cmn}}$$

and the Lorentz-Lorentz formula for the internal field. The introduction of an internal field implies the absence of resonances, which is not fulfilled in a wide range, therefore formula (13) is problematic. The frequencies $\omega_{cmn}$ characterize the relaxation times of the levels. If the concentrations $N_m$ of atoms of the $m$ variety (squares of plasma frequencies $\omega_{pm}^2$) are small, i.e. the sum is small compared to unity, (13) can be decomposed into a small parameter:

$$\varepsilon(\omega) \approx 1 + \sum_{n,m=1} \frac{\omega_{pm}^2}{\omega_{mn}^2 - \omega^2 + i\omega_{cmn}\omega}. \tag{14}$$

This formula is derived from the Lorentz oscillator model [39] and is often used, although it is strictly valid for a rarefied gas of oscillators with several resonant frequencies $\omega_{mn}$. Next, we use it, since formula (13) leads to inadequate results at resonances (small distances). It can be used in the low-frequency range at high $d$. If there are atoms of only one kind, then $m=1$. If there is only one resonant frequency, then $n=1$. The values $\omega_{pm}^2$ characterize the oscillator forces calculated from solving a quantum mechanics problem. If $\omega = \omega_{mn}$ for zero CF $\varepsilon(\omega_{mn}) \approx -2 < 0$, and formula (13) cannot be used in this case, as in the case of equality of the sum to three ($\varepsilon = \infty$), since it is obtained in the approximation of a small sum. In real media with a large number of frequencies, significant losses, and small oscillator forces, for most oscillations, the real part of the DP $\varepsilon'$ does not go through zero. Such a transition usually takes place in metals. Consideration of media with dispersion (14) is of interest [47]. Note that for the region significantly lower than the resonant frequencies, an "optical" or transparent part of the DP is obtained, determined by the low-frequency polarization of the substance:

$$\varepsilon_L = 1 + \sum_{n,m=1} \frac{\omega_{pm}^2}{\omega_{mn}^2}.$$

The squares of plasma frequencies (PF) determine the concentrations of atoms and usually lie in the UV range. For metals, there are free electrons. In the model, this means a zero resonant frequency (no coupling), which characterizes additional electronic susceptibility.

$$\chi_e = -\frac{\omega_p^2}{\omega^2 - i\omega_c\omega},$$

determined by the PF and CF for conduction electrons. For them, the resonant frequency is zero because they are free and not bound to atoms. Note that from (14) it is also possible to obtain the Debye dispersion law in the limit for absolutely rigid dipoles (high transition frequencies) with orientational polarization [42]. The considered models allow us to accurately describe the real media, if we take into account a sufficient number of members. Actually condensing atomic spectra have many (infinitely many) terms. Additional spectral terms arise for polyatomic systems and molecules, so it is easier to determine DP through the absorption spectrum [3, 48], which can be experimentally measured in a wide range. However, this is inconvenient for analytical and numerical calculations. Taking into account a sufficient number of terms allows us to build an adequate model of the dispersion forces. The transition to a complex frequency means the dependence

$$\varepsilon(k) = 1 + \chi_e(k) + \sum_{n,m=1} \frac{k_{pm}^2}{k_{mn}^2 + k^2 + k_{cmn}k}, \quad (15)$$

where the corresponding wave numbers are entered. Also $\chi_e(k) = k_p^2/(k^2 + k_c k)$. This value has poles at $k = -k_c$ and at $k = 0$. To avoid the latter, the Drude-Smith model can be used in Ref. [49,50]. In finite structures, a free electron cannot escape to infinity from an atom, i.e. it can be approximately characterized by a very small coupling constant $k_s^2$ related to size, and susceptibility can be introduced $\chi_e(k) = k_p^2/(k_s^2 + k^2 - k_c k)$. You can take it $k_s \sim k_c$, but with a very large thickness $k_s \sim 1/t$. It's important that $\varepsilon(\infty) = 1$. This means that for $k \to \infty$ we have $\varepsilon \to 1$, and in formulas (2), (3) $K \to K_0$, and $f_{e,h}(\kappa,k) \to \infty$, providing, along with a large factor $\exp(2K_0 d)$, the convergence of the integral (4). Other DP models are possible, including accounting for the internal field, for example, according to the Onsager formula [39].

To numerically calculate the integrals (5) or (7), we turn to the polar coordinates $\kappa = \chi\cos(\theta)$, $k = \chi\sin(\theta)$, $K_0 = \chi$. At the point $\theta = 0$ we have $k = 0$ and $\varepsilon(0)$ – the low-frequency DP value. At $\theta \to 0$ the DP's commitment to $\varepsilon(0)$ provides a significant contribution to strength. At all other points $\theta > 0$, DP tends to unity at $\chi \to \infty$. Therefore, the angle integral is divided into two intervals $(0, \theta_0)$ and $(\theta_0, \pi/2)$. In the first case, we perform careful integration by angle, and if the angle is small, then $K = \chi\sqrt{1 + \sin^2(\theta)\varepsilon(\chi,\theta)} \approx \chi\sqrt{1 + \varepsilon(0)}$. The ratio (2), (3) for large values $\chi$ is written as $f_{e,h}(\chi,\theta) = \exp(2\chi d)\varphi_{e,h}(\chi,\theta) - 1 \approx \exp(2\chi d)\varphi_{e,h}(\chi,0)$. We select the integration areas $0 < \chi < \chi_0$ and $\chi_0 < \chi < \infty$. For the second region, considering a large value of $\chi_0$, we have independent of $\chi$ functions

$$\varphi_h(\chi,0) = \left[\frac{2\sqrt{1+\varepsilon(0)} + 2 + \varepsilon(0)}{\varepsilon(0)}\right]^2, \quad (16)$$

$$\varphi_e(\chi,0) = \left[\frac{1 + 2\varepsilon(0)\sqrt{1+\varepsilon(0)} + \varepsilon(0) + \varepsilon^2(0)}{1 + \varepsilon(0) - \varepsilon^2(0)}\right]^2, \tag{17}$$

and the result for the integral of the remainder is

$$\int_{\chi_0}^{\infty} \chi^2 \exp(-2\chi d)\left(\frac{1}{\varphi_e(\chi,0)} + \frac{1}{\varphi_h(\chi,0)}\right) d\chi =$$

$$= \exp(-2d\chi_0)\left(\frac{1}{\varphi_e(\chi_0,0)} + \frac{1}{\varphi_h(\chi_0,0)}\right)\left(\frac{\chi_0^2}{2d} + \frac{\chi_0}{2d^2} + \frac{1}{4d^3}\right).$$

This result allows us to choose $\chi_0$ so that the integral of the remainder is significantly less than the integral in the domain $0 < \chi < \chi_0$. For an area $(\theta_0, \pi/2)$, it is enough to take several points of integration along the angle 500 were used in the calculations. The integral can even be calculated approximately by the mean value theorem at a point $\tilde{\theta} = (\theta_0 + \pi/2)/2$. Then $K = \tilde{K}(\chi) = \chi\sqrt{1 + \varepsilon(\chi,\tilde{\theta})\sin^2(\tilde{\theta})}$, and for the integral over $\chi$ we have

$$\int_0^{\infty}\left(\frac{1}{\exp(2\chi d)\varphi_e(\chi,\tilde{\theta})-1} + \frac{1}{\exp(2\chi d)\varphi_h(\chi,\tilde{\theta})-1}\right)\chi^2 d\chi =$$

$$\int_0^{\chi_0}\left(\frac{1}{\exp(2\chi d)\varphi_e(\chi,\tilde{\theta})-1} + \frac{1}{\exp(2\chi d)\varphi_h(\chi,\tilde{\theta})-1}\right)\chi^2 d\chi + \tag{18}$$

$$+ \exp(-2d\chi_0)\left(\frac{1}{\varphi_e(\chi_0,\tilde{\theta})} + \frac{1}{\varphi_h(\chi_0,\tilde{\theta})}\right)\left(\frac{\chi_0^2}{2d} + \frac{\chi_0}{2d^2} + \frac{1}{4d^3}\right)$$

The value $\chi_0$ should be selected from the conditions $\chi_0^2 \gg k_{mn}^2/\sin^2(\theta_0)$, $\chi_0^2 \gg k_{pn}^2/\sin^2(\theta_0)$ $k_{pm}^2$. The values $k_{pn}^2$ are related to the concentration of atoms, and the wavelengths $\lambda_{pn} = 2\pi/k_{pn}$ usually correspond to the UV range. The transition frequencies may be higher and correspond to energies of the order of several eV. Therefore, the minimum wavelengths $\lambda_{min}$ are of the order of several tens of nm, and the magnitude $\chi_0 > 2\pi/\lambda_{min}$ is of the order of 0.1 (1/nm). This upper limit makes it possible to calculate integrals very accurately.

Consider the behavior of the force at large distances $d$. Making the substitution $\kappa = x/d$, $k = y/d$, we bring (5) to the form

$$P(d) = -\frac{\hbar c}{2\pi^2 d^4}\int_0^{\infty} x dx \int_0^{\infty} \sqrt{x^2 + y^2}\left(\frac{1}{f_e(x,y,d)} + \frac{1}{f_h(x,y,d)}\right) dy. \tag{19}$$

For large $d$ the function

$$f_h(x,y,d) = \exp\left(2\sqrt{x^2+y^2}\right)\left[\frac{2\sqrt{x^2+y^2}\sqrt{x^2+\varepsilon(x,y)y^2} + 2x^2 + (\varepsilon(x,y)+1)y^2}{\varepsilon(x,y)y^2 - y^2}\right]^2 - 1$$

does not depend on this distance $d$. The function $f_e(x,y,d)$ is also independent, so we have $P(d) \sim 1/d^4$. Exponentially small additions provide corrections to this dependence. By making the substitution $u = 2K_0 d = 2d\sqrt{\kappa^2 + v^2/d^2}$, $udu/(2d)^2 = \kappa d\kappa$, $v = kd$, we obtain the integrals

$$P(d) = -\frac{\hbar c}{16\pi^2 d^4} \int_0^\infty \int_{2v}^\infty \left( \frac{\varphi_e^{-1}(u,v,d)}{1-\varphi_e^{-1}(u,v,d)\exp(-u)} + \frac{\varphi_h^{-1}(u,v,d)}{1-\varphi_h^{-1}(u,v,d)\exp(-u)} \right) \exp(-u) u^2 dv du.$$

With a large distance $d$, the functions $\varphi_{e,h}$ cease to depend on it, and the integral over $u$ can be approximately calculated by integrating by parts and discarding the small remainder. Denoting the parenthesis as $\Phi(u,v)$, we get

$$P(d) \approx -\frac{\hbar c}{16\pi^2 d^4} \int_0^\infty dv \exp(-2v) \left[ v^2 \Phi(2v,v) + \delta_u\left(u^2\Phi(u,v)\right)_{u=2v} + \delta_u^2\left(u^2\Phi(u,v)\right)_{u=2v} \right].$$

The first term in the square bracket has a second-order zero at zero, so the integral of it can also be approximated by integration by parts three times. As a result, we have a nonintegrative term $\Phi(0,0)/4$ and a contribution to the integral of $[4v\Phi'(2v,v)+v^2\Phi''(2v,v)]/8$. A stroke means differentiation by the first variable. The second and third terms are equal to $2v\Phi(2v,v)+v^2\Phi'(2v,v)$ and $2\Phi(2v,v)+4v\Phi'(2v,v)+v^2\Phi''(2v,v)$. They also have first- and second-order zeros, so the process can be continued. As a result, it is possible to obtain the decomposition of the derivatives of the function $\Phi$ at zero. If $\varepsilon \to \infty$, then $\varphi_{e,h} \to 1$, and after substitution $K_0 = pk$, $\kappa = k\sqrt{p^2-1}$, we have the Casimir result $P(d) = -\hbar c \pi^2/(240 d^4)$.

Consider the following Lifshitz problem from (5):

$$P(d) = -\frac{\hbar c}{2\pi^2} \int_1^\infty p^2 dp \int_0^\infty k^3 \left( \frac{1}{\exp(2pkd)S^2(p,1)-1} + \frac{1}{\exp(2pkd)S^2(p,\varepsilon)-1} \right) dk,$$

$s(p) = \sqrt{p^2-1+\varepsilon}$, $S(p,\varepsilon) = (s(p)+\varepsilon(k)p)/(s(p)-\varepsilon(k)p)$. Making a substitution $k = v/d$, we have

$$P(d) = -\frac{\hbar c}{2\pi^2 d^4} \int_1^\infty p^2 dp \int_0^\infty v^3 \left( [\exp(2pv)S^2(p,1)-1]^{-1} + [\exp(2pv)S^2(p,\varepsilon)-1]^{-1} \right) dv.$$

Assuming that the main contribution takes place at $p \approx 1$ and counting $s(1) = \sqrt{\varepsilon}$, we have

$$P(d) \approx -\frac{\hbar c}{2\pi^2 d^4} \int_1^\infty p^2 dp \int_0^\infty v^3 \left( [\exp(2pv)S^2(1,1)-1]^{-1} + [\exp(2pv)S^2(p,\varepsilon)]^{-1} \right) dv,$$

$S(1,\varepsilon) = (\sqrt{\varepsilon(v)}+\varepsilon(v))/(\sqrt{\varepsilon(v)}-\varepsilon(v))$. Ignoring the units, we find

$$P(d) \approx -\frac{\hbar c}{2\pi^2 d^4} \int_1^\infty p^2 dp \int_0^\infty v^3 \times \left( \exp(-2pv)\left(\frac{\sqrt{\varepsilon(v)}-1}{\sqrt{\varepsilon(v)}+1}\right)^2 + \right.$$
$$\left. + \exp(-2pv)\left(\frac{\sqrt{\varepsilon(v)}-\varepsilon(v)}{\sqrt{\varepsilon(v)}+\varepsilon(v)}\right)^2 \right) dv.$$

Calculating the integrals with respect to $p$, we obtain

$$\int_1^\infty p^2 \exp(-2pv)dp = \exp(-2v)\left[\frac{1}{2v}+\frac{1}{2v^2}+\frac{1}{4v^3}\right].$$

The result can be easily obtained if a low-frequency DP $\varepsilon(v) \approx \varepsilon(0)$ is used in the entire range where dissipation occurs:

$$P(d) \approx -\frac{3\hbar c}{8\pi^2 d^4}\left(\frac{1-\sqrt{\varepsilon(0)}}{1+\sqrt{\varepsilon(0)}}\right)^2. \tag{17}$$

This requires that the value $v = \xi d/c$ be small, i.e. $d < c/\xi_{max}$. If the transition frequencies lie in the UV region and are on the order of $10^{16}$ Hz, this means distances $d < 30$ nm. For diamond $\varepsilon(0) = 5.6$, and we obtain a force 0.448 times less than in the Casimir model. Assuming as in [3] $s(p) = p$, we find after the substitution $2pv = x$

$$P(d) \approx -\frac{\hbar c}{16\pi^2 d^3} \int_0^\infty \int_{\xi d/c}^\infty \frac{x^2}{\exp(x)\left(\frac{1+\varepsilon(v)}{1-\varepsilon(v)}\right)^2 - 1} dv dx.$$

There is a lower limit $v = \xi d/c$ in this formula, so it coincides with the formula from [3] (in the latter, the lower limit is taken as zero), i.e. it gives a dependence $1/d^3$. However, this is a transitional dependence from large to small distances. For very small $d$, the force is finite. Also assuming $\varepsilon(v) \approx \varepsilon(0)$ we find

$$P(d) \approx -\frac{\hbar}{16\pi^2 d^3}\left(\frac{1-\varepsilon(0)}{1+\varepsilon(0)}\right)^2 \int_0^\infty \int_{2\xi d/c}^\infty \frac{x^2 \exp(-x)}{1-\exp(-x)\left(\frac{1-\varepsilon(0)}{1+\varepsilon(0)}\right)^2} d\xi dx,$$

$$P(d) \approx -\frac{\hbar}{16\pi^2 d^3}\left(\frac{1-\varepsilon(0)}{1+\varepsilon(0)}\right)^2 \int_0^\infty d\xi \int_{2\xi d/c}^\infty x^2 \left(\exp(-x)+\exp(-2x)\left(\frac{1-\varepsilon(0)}{1+\varepsilon(0)}\right)^2\right) dx.$$

The integral over $x$ has the value

$$\int_{2\xi d/c}^{\infty} x^2 \left[ \exp(-x) + \exp(-2x)\left(\frac{1-\varepsilon(0)}{1+\varepsilon(0)}\right)^2 \right] dx =$$
$$= \exp(-2\xi d/c)\left[(2\xi d/c)^2 + 2(2\xi d/c) + 2\right] +$$
$$+ \exp(-4\xi d/c)\left[\frac{(2\xi d/c)^2}{2} + \frac{(2\xi d/c)}{2} + \frac{1}{4}\right]\left(\frac{1-\varepsilon(0)}{1+\varepsilon(0)}\right)^2$$

Now the integral over $\xi$ is also easily calculated, which gives terms proportional to $(c/d)^\nu$, $\nu = 0,1,2$, i.e., in addition to the term $1/d^3$, there are terms with $1/d^4$ and $1/d^5$ by taking into account the lower limit. We do not provide the final result.

The Van Kampen method with functions of type (9)–(12) does not formally allow calculating the result for very small $d$. Indeed, it is based on the principle (or theorem) of the argument and requires the vanishing of the integral on the large right semicircle of the complex plane $\xi$ (or $k$). This provides a large multiplier $\exp(2K_0 d) = \exp\left(2\sqrt{\kappa^2 + k^2}\, d\right)$ in the denominator. However, when $d = 0$ it is equal to one and does not ensure convergence. Accordingly, it cannot be decomposed in $d$, and for small $d$, the upper limit should be increased with the condition $\chi_0 > 2\pi/d_{\min}$. So, for $d = 1$ nm we have $\chi_0 > 2\pi$ (1/nm). Since at high frequencies $\varepsilon \to 1$, $\varepsilon(k) - 1 \approx k_{\max}^2/k^2$, then $K = K_0\left(1 + k_{\max}^2/(2k^2 K_0)\right) \to K_0$, $K_0 \to \infty$, and at $d \ll \sqrt{\kappa_{\max}^2 + k_{\max}^2}/2$ we get

$$\frac{1}{f_h(\kappa,k,0)} = \frac{k_{\max}^4}{16(\kappa^2+k^2)^2 - k_{\max}^4} \to 0,$$

$$\frac{1}{f_e(\kappa,k,0)} = \frac{(2\kappa^2+k^2)^2 k_{\max}^4}{4K_0^4 k^4 - (2\kappa^2+k^2)^2 k_{\max}^4} \to 0.$$

However, at $d=0$, the result (5) does not exist. Indeed, the remainder of the integral from $K_0 f_h^{-1}(\kappa,k,0)$ at high frequencies is

$$\frac{k_{\max}^4}{16}\int_0^\infty \kappa d\kappa \int_{k_{\max}}^\infty \frac{K_0 dk}{K_0^4 - k_{\max}^4/16} \approx \frac{k_{\max}^4}{16}\int_0^\infty d\kappa \left(\frac{1}{\kappa} - \frac{k_{\max}}{\kappa\sqrt{\kappa^2 + k_{\max}^2}}\right).$$

It is logarithmically diverging. The remainder for $f_e^{-1}(\kappa,k,0)$ also diverges. In principle, integrals can be calculated for any small but finite $d$. But as $d$ decreases, the upper limit should be increased proportionally $1/d$.

Consider the case of a dielectric with DP $\tilde{\varepsilon}(\omega)$ between the plates. In this case, instead $K_0$, we should take in exponent $\tilde{K} = \sqrt{\kappa^2 + \tilde{\varepsilon}k^2}$ and functions of the form [45]

$$f_h(\kappa,k,d) = \exp(2\tilde{K}d)\left[\frac{K(\tilde{K}+K_0)\coth(Kt)+K^2+\tilde{K}K_0}{K(\tilde{K}-K_0)\coth(Kt)+K^2-\tilde{K}K_0}\right]^2 - 1, \qquad (18)$$

$$f_e(\kappa,k,d) = \exp(2\tilde{K}d)\left[\frac{\varepsilon K(\tilde{K}+\tilde{\varepsilon}K_0)\coth(Kt)+\tilde{\varepsilon}\tilde{K}K_0+\varepsilon^2 K^2}{\varepsilon K(\tilde{K}-\tilde{\varepsilon}K_0)\coth(Kt)+\tilde{\varepsilon}\tilde{K}K_0-\varepsilon^2 K^2}\right]^2 - 1. \qquad (19)$$

For thick layers ($\coth(Kt)=1$) and for a thin film of thickness $d$ with DP $\tilde{\varepsilon}$ between them, we obtain the result [24] corresponding to the result of [23]:

$$P(d) \approx -\frac{\hbar}{16\pi^2 d^3}\int_0^\infty d\xi \int_{2\xi d/c}^\infty \left\{\exp(x)\left(\frac{[\varepsilon(\xi)+\tilde{\varepsilon}(\xi)]^2}{[\varepsilon(\xi)-\tilde{\varepsilon}(\xi)]^2}\right)-1\right\}x^2 dx.$$

The absence of a film ($\tilde{\varepsilon}=1$) corresponds to the Lifshitz result (3.1L) at $\varepsilon_1=\varepsilon_2=\varepsilon$ and the zero lower limit. In the case of $t=0$ from (18), (19) we obtain

$$f_h(\kappa,k,d) = \exp(2\tilde{K}d)\frac{(\tilde{K}+K_0)^2}{(\tilde{K}-K_0)^2} - 1,$$

$$f_e(\kappa,k,d) = \exp(2\tilde{K}d)\frac{(\tilde{K}+\tilde{\varepsilon}K_0)^2}{(\tilde{K}-\tilde{\varepsilon}K_0)^2} - 1.$$

The result (5) with these functions corresponds to the external Casimir pressure on a film of thickness $d$ with DP $\tilde{\varepsilon}$ located in a vacuum. For a film with a very large DP $\tilde{\varepsilon}=1+k_{max}^2/k^2$ (for dense plasma at $k_{max}/k \gg 1$) we have

$$(K+K_0)/(K-K_0) = \left(\sqrt{p^2-1+\varepsilon}+p\right)/\left(\sqrt{p^2-1+\varepsilon}-p\right) \approx 1$$

in the area where the main contribution to the force takes place (small $p$), or $f_{e,h}(\kappa,k,d) \approx \exp\left(2kd\sqrt{p^2+k_{max}^2/k^2}\right)-1$. The unit can be ignored if $k < k_{max}$ and $2k_{max}d > 1$. Making a substitution $y = 2kd\sqrt{p^2+\tilde{\varepsilon}(k)-1}$ or $y^2 = 4d^2(k^2 p^2 + k_{max}^2)$, we get

$$P(d) = -\frac{\hbar c}{16\pi^2 d^4}\int_1^\infty p^{-2}dp \int_{2dk_{max}}^\infty \frac{y(y^2-4d^2 k_{max}^2)}{\exp(y)-1}dy.$$

For a thin layer $dk_{max} \to 0$, we obtain the Casimir pressure $P(d) = -\hbar c\pi^2/(240 d^4)$. However, the assumption $k_{max}$ is large, so going to the limit is impossible. The integral should be calculated strictly. If we replace the lower limit with zero, the Bose-Einstein integrals are calculated and we get a correction

$$P(d) = -\frac{\hbar c}{24\pi^2 d^4}\left(1-10d^2 k_{max}^2\right).$$

It says that as $d$ increases, the pressure decreases faster than $1/d^4$, and at a certain distance it can disappear or even change its sign. However, the formula is approximate, and a rigorous result requires numerical integration. This pressure can be explained by the van der Waals attraction of molecules. The result cannot be used at molecular distances. It should be considered at $d>1$ nm, and at shorter distances the force is finite.

In the case of a large number of layers, the characteristic equation is obtained by the transmission matrix method [46]. For the Lifshitz problem, it is easier to obtain the characteristic equation by transforming the impedance. So, the normalized impedance of E-mode is $\rho_e = \sqrt{\kappa^2 + k^2\varepsilon}/(k\varepsilon)$, and the impedance H- mode is $\rho_h = k/K$. For an empty space (slot) $\rho_{0e} = K_0/k$, $\rho_{0h} = k/K_0$. The impedances are transformed by the slot to the impedances

$$Z = \rho_{0e,h} \frac{\rho_{e,h} + i\rho_{0e,h}\tan(k_{z0}d)}{\rho_{0e,h} + i\rho_{e,h}\tan(k_{z0}d)} =$$
$$= \rho_{0e,h} \frac{\rho_{e,h} + \rho_{0e,h}\tanh(K_0 d)}{\rho_{0e,h} + \rho_{e,h}\tanh(K_0 d)}.$$

Here $k_{z0} = -iK_0$. For resonance it is necessary $Z = -\rho_{e,h}$, from where we get the equation

$$\tilde{f}_{e,h}(\kappa,k,d) = \rho_{0e,h} \frac{\rho_{e,h} + \rho_{0e,h}\tanh(K_0 d)}{\rho_{0e,h} + \rho_{e,h}\tanh(K_0 d)} + \rho_{e,h} = 0. \quad (20)$$

We have $\tilde{f}_{e,h}(\kappa,k,\infty) = \rho_{e,h} + \rho_{0e,h}$, $\tilde{f}_{e,h}(\kappa,k,0) = 2\rho_{e,h}$. However, these functions correspond to functions (9) and (10) up to multipliers. Replacing the hyperbolic tangent by $(\exp(2K_0 d)-1)/(\exp(2K_0 d)+1)$, we find

$$\tilde{f}_{e,h}(\kappa,k,d) = \frac{(\rho_{0e,h} + \rho_{e,h})^2 \exp(2K_0 d) - (\rho_{0e,h} - \rho_{e,h})^2}{(\rho_{e,h} + \rho_{0e,h})\exp(2K_0 d) + \rho_{0e,h} - \rho_{e,h}}. \quad (21)$$

Integral (7) with function (21) diverges for any finite or even infinite $d$. According to the principle of the argument, it is determined with precision to a certain value associated with the infinite vacuum energy [29]. The value as a difference

$$\frac{1}{\bar{f}_{e,h}(\kappa,k,d)} = \frac{1}{\tilde{f}_{e,h}(\kappa,k,d)} - \frac{1}{\tilde{f}_{e,h}(\kappa,k,\infty)} =$$
$$= \frac{2\rho_{0e,h}(\rho_{0e,h} - \rho_{e,h})}{(\rho_{0e,h} + \rho_{e,h})^2 \exp(2K_0 d) - (\rho_{0e,h} - \rho_{e,h})^2} \quad (22)$$

at large $d$ vanishes, i.e. integral (7) exists with it at such distances and describes (up to a factor) the force at large distances. To match the Lifshitz problem $\tilde{\tilde{f}}_{e,h} = \bar{f}_{e,h}$, we should take the function $\tilde{\tilde{f}}_{e,h}(\kappa,k,d) = 2(\rho_{0e,h} - \rho_{e,h})\bar{f}_{e,h}(\kappa,k,d)/\rho_{0e,h}$. Assuming $\varepsilon \to \infty$ ($\rho_{e,h} \to 0$), we find the correspondence of this function to the Casimir problem $\tilde{\tilde{f}}_{e,h}(\kappa,k,d) = \exp(2K_0 d) - 1$. In particular, for the Lifshitz problem with zero gap, we obtain

$$f_{e,h} = \frac{(\rho_{0e,h} + \rho_{e,h})^2 - (\rho_{0e,h} - \rho_{e,h})^2}{(\rho_{0e,h} - \rho_{e,h})^2}.$$

At high frequencies $\varepsilon(k) = 1 + k_{max}^2/k^2$, and we have $K \approx K_0 + k_{max}^2/(2K_0)$, where $K_0$ is a large value. Therefore $f_e \approx 4k^2(1 + k_{max}^2/k^2)(K_0^2 + k_{max}^2)/k_{max}^4 \approx 4k^2 K_0^2/k_{max}^4$, and similarly $f_h \approx 16 K_0^3 (K_0 + k_{max}^2/(2K_0))/k_{max}^4 \approx 16 K_0^4/k_{max}^4$. These values are large, but they do not ensure convergence of the integrals. Indeed, consider the integral for H-modes:

$$\int_0^\infty \kappa d\kappa \int_{k \gg k_{max}}^\infty \frac{K_0}{f_h(\kappa,k)} dk \approx k_{max}^4 \int_0^\infty \kappa d\kappa \int_{k \gg k_{max}}^\infty \frac{1}{16 K_0^3}\left(1 - \frac{2k_{max}^2}{4K_0^2}\right) dk.$$

Replacing the parenthesis with unity, we get a logarithmically divergent integral

$$\frac{k_{max}^2}{16} \int_{k \gg k_{max}}^\infty \frac{dk}{k}.$$

Accordingly, the Van Kampen method does not allow calculating the force with an infinitesimal (zero) gap. Consider the corresponding equation $f_h(\kappa, k_0) = 0$, which takes the form $k = \ln(\pm \gamma(k,p))/(pd)$, because

$$\exp(K_0 d) = \exp(kpd) = \pm\left(\frac{K - K_0}{K + K_0}\right) = \pm \gamma(p,k).$$

All branches of the logarithm should be taken into account here. It follows that for small $d$, the zeros are shifted to the high frequency range. The motion of zeros in the complex plane is shown in [10] (Fig. 3). This applies to low frequencies and short distances. For high frequencies $K \approx K_0$, $\gamma(p,k) = 0$, and the value $k = \ln(\pm \gamma)/(pd)$ at $d \to 0$ becomes indeterminate. At low $d$, all frequencies become large, and the plates are transparent to them. This suggests that the value $P(0)$ is finite. Indeed, the infinite attraction of the two plates would release infinite energy, which is physically absurd. Although, on the other hand, the continuum model no longer holds in this case. The frequencies can be found by solving the equation ( $n = 0, \pm 1 \pm ....$ )

$$k = \frac{i\omega}{c} = \frac{1}{pd} \ln\left(\pm \frac{\sqrt{p^2 - 1 + \varepsilon(k)} - p}{\sqrt{p^2 - 1 + \varepsilon(k)} + p}\right) + \frac{2i\pi n}{pd}.$$

Hence, for the real parts we have $\omega_n' = \pm 2\pi n/(pd)$, $n = 1,2,...$, and for the imaginary parts

$$\omega_n'' = \pm \frac{1}{pd} \ln\left(\frac{\sqrt{p^2 - 1 + \varepsilon(k)} - p}{\sqrt{p^2 - 1 + \varepsilon(k)} + p}\right).$$

Thus, at small distances, all frequencies are shifted to an infinite region. Infinite frequencies are not perturbed by the dielectric, so the contribution to the perturbation energy is zero or at least finite. On the other hand, equation (20) $\tilde{f}_{e,h}(\kappa,k,0)=0$ implies $\rho_{e,h}=0$, or $\rho_e = \sqrt{p^2 -1+\varepsilon(k)}=0$. In the case $\varepsilon(k)>1$, the equation has no zeros in the finite domain. The equation $\rho_h = 0$ has zero at $p=\infty$. Similarly, for two plates at $d=0$, all resonant frequencies are shifted to infinity, so the force density is finite.

For the Lifshitz problem $K_0 > k_{max}$, and we have

$$\frac{1}{f_h(\kappa,k)} \approx \frac{k_{max}^4/(4K_0^2)}{\exp(2K_0 d)(4K_0^2 + 2k_{max}^2)^2 - k_{max}^4/(4K_0^2)} \approx$$
$$\approx \frac{k_{max}^4}{16K_0^4 \exp(2K_0 d)}\left(1+\frac{k_{max}^4}{16K_0^4 \exp(2K_0 d)}\right)$$

Convergence will be if $\exp(2K_0 d) \geq (2K_0 d)^\nu$. We have the equation $\exp(x)=x^\nu$ and its root $x_0 = 2K_0 d$, $x_0 \approx 2$ for $\nu=3$. We obtain the convergence condition of the integral for small $d$: $\kappa^2 + k^2 \gg x_0^2/4d^2 \approx 1/d^2$. It is the same for $f_e(\kappa,k_0)$. The $\Theta(\tilde{\omega}_n)$ should be used in (1) instead of $\hbar\tilde{\omega}_n/2$. Note that $\Theta$ is an even function of frequency (positive for negative frequencies). Also, the functions $f_{e,h}(p,\omega)$ are even. Now (7) should be taken into account as

$$P(d,T) = -\frac{\hbar}{2\pi^2 c^3}\int_1^\infty p^2 dp \int \omega^3 \coth\left(\frac{\hbar\omega}{2k_B T}\right)\left(\frac{1}{f_e(p,\omega)}+\frac{1}{f_h(p,\omega)}\right)d\omega,$$

or as

$$P(d,T) = -\frac{\hbar c}{2\pi^2}\int_1^\infty p^2 dp \int_0^\infty \cot\left(\frac{\hbar kc}{2k_B T}\right)\left(\frac{1}{f_e(p,k)}+\frac{1}{f_h(p,k)}\right)k^3 dk, \qquad (23)$$

which coincides with the Lifshitz formula for the finite temperature [3]. In the case of high temperatures $k_B T \gg \hbar\omega$ for frequencies under consideration (hot plasma) will be

$$P(d,T) = -\frac{k_B T}{\pi^2}\int_1^\infty p^2 dp \int_0^\infty \left(\frac{1}{f_e(p,k)}+\frac{1}{f_h(p,k)}\right)k^2 dk.$$

Let find a correction to formula (7) at a small finite (on the order of room) temperature using decomposition $\coth(x) \approx (1+\exp(-2x))^2 \approx 1+2\exp(-2x)$ at large x. We have

$$P(d,T) = P(d,0) - \frac{\hbar}{\pi^2 c^3}\int_1^\infty p^2 dp \int \omega^3 \exp\left(-\frac{\hbar\omega}{k_B T}\right)\left(\frac{1}{f_e(p,\omega)}+\frac{1}{f_h(p,\omega)}\right)d\omega.$$

Integral (23) has poles $k_n = 2n\pi k_B T/(\hbar c) = i\omega_n/c$. It can be calculated by replacing, as usual, the integral by the sum of the Matsubara frequencies [3,10] $\omega_n = i\xi_n = ick_n$, i.e. by taking half-residuals at $\omega_n$ and a quarter at $\omega_0 = 0$:

$$P(d) = -\frac{8\pi^2 (k_B T)^4}{c^3 \hbar^3} \sum_{n=0}^{\infty} \frac{1}{1+\delta_{n0}} \int_1^{\infty} \left( \frac{1}{f_e(x/n,\omega_n)} + \frac{1}{f_h(x/n,\omega_n)} \right) x^2 dx. \qquad (24)$$

Here, the quantities $K(x/n,\omega_n) = k_n \sqrt{x^2/n^2 + \varepsilon - 1}$ are in the functions $f_{e,h}$. Since the integrative function in (23) is even in $k$, the integral can be extended to the entire axis and the integration contour can be taken as shown in Fig. 3.7 of Ref. [10]. Since the frequencies $\omega_{mnl} = \omega'_{mnl} \pm i\omega''_{mnl}$ are located in the right half-plane $\omega$ and are complex conjugate, such integration yields a doubled sum over the positive frequencies. Since the poles are simple, enclosing them with small neighborhoods, integral (23) can be calculated numerically in the sense of the main value. In formula (24), $n=0$ corresponds to $f_e(\infty,\omega_n) = \infty$, and this contribution, as it is easy to see, is absent. Thus, the Van Kampen method allows considering an arbitrary number of layers by constructing a characteristic equation, as well as taking temperature into account. It also allows you to insert conductive, for example, graphene sheets into the layers. In the simplest case of weighted sheets with normalized conductivity $\zeta$, we have

$$f_{e,h} = \frac{(\varsigma \rho_{e,h})^2}{(2+\varsigma \rho_{e,h})^2 \exp(2K_0 d) - (\varsigma \rho_{e,h})^2}.$$

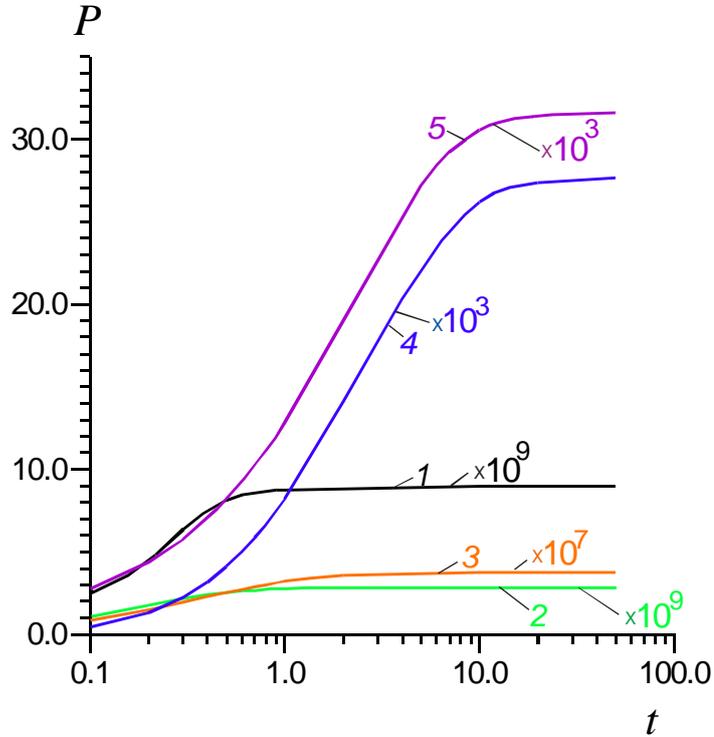

Fig. 2. Casimir pressure $P=F/L^2$ (N/m$^2$) between two dielectric layers depending on their thickness $t$ (nm) at different distances $d$ (nm): $d=0.01$ (curve 1); 0.1 (2); 1 (3); 10 (4,5). Curves 1–4 are plotted in the absence of conductivity ($\omega_p = 0$), curve 5 – in the presence of it

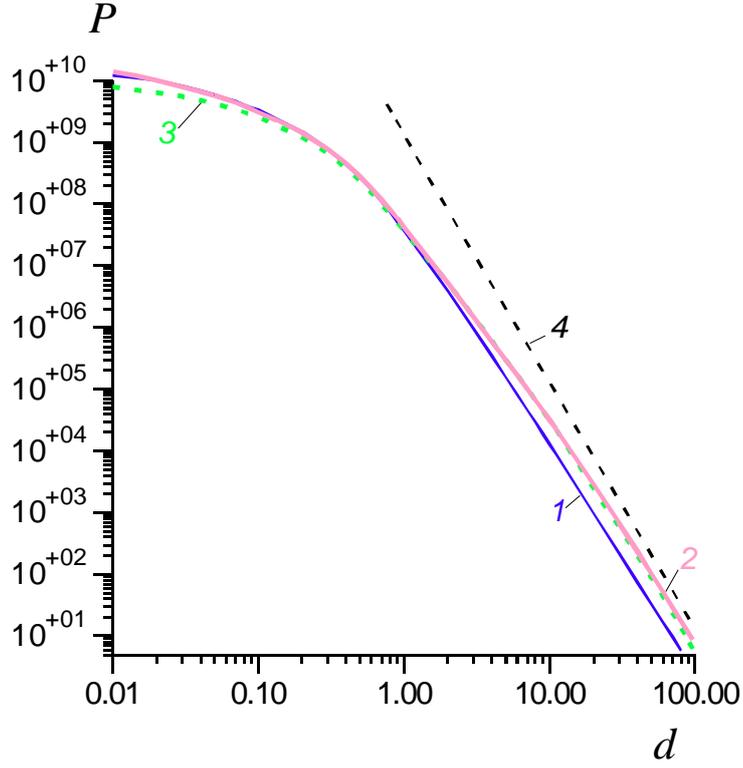

Fig. 3. Casimir pressure P (N/m$^2$) between two dielectric layers depending on the distance d (nm) at different thicknesses $t$ (nm): $t = 1$ (curve 1); 50 (2.3). Curves 1, 2 are constructed taking into account the conductivity, curve 3 at $\omega_p = 0$. 4 – the Casimir result

## 4. Numerical results

The model (15) is used for dielectric plates: $m=1$, $n=6$, $k_p = 0.05$, $k_{pn} = 0.05$, $n=1,\ldots,6$, $k_{r1} = 0.01$, $k_{r2} = 0.02$, $k_{r3} = 0.03$, $k_{r4} = 0.04$, $k_{r5} = 0.05$, $k_{r6} = 0.08$, $k_{c0} = k_{cn} = 10^{-6}$ (everything is in reverse nm). The numerical results are shown in Fig. 2 and 3. The dependence of pressure $P$ on the thickness of the plates at different distances is shown in Fig. 2. Curves 1–4 correspond to the absence of conductivity: $k_p = 0$, $k_{pn} = 0.05$, curve 5 is constructed at $k_p = 0.05$. All curves are saturated at thicknesses of the order of 10 nm, so measurements with such plates give the Lifshitz configuration force. The dependence of pressure $P$ on distance at different plate thicknesses is shown in Fig. 3. One can see the difference at short distances from the law $1/d^4$, which is carried out at long distances. This difference is already strongly evident at $d < 10$ nm. When $d \to 0$, the pressure tends to the finished value. The case of the absence of conductivity is also considered there ($k_p = 0$, curve 3). At $t \sim 1$ nm or less, the results for the van der Waals force are completely obtained by the method of density functional theory and correspond to the above. Integrals of type (4) were calculated by replacing $\kappa = \chi \cos(\theta)$, $k = \chi \sin(\theta)$ using 600 points of integration along the angle and 5000 points of integration along $\chi$, and the region $\chi$ was divided into 6 subdomains with simultaneous integration into them.

The lower area matched $0 < \chi < k_{c0}$. The upper area corresponded to $k_{r6} < \chi < k_{max}$ where $k_{max} = 1 + 10k_{r6} + 1/d$. The choice of the specified number of points guaranteed an accuracy of three decimal places.

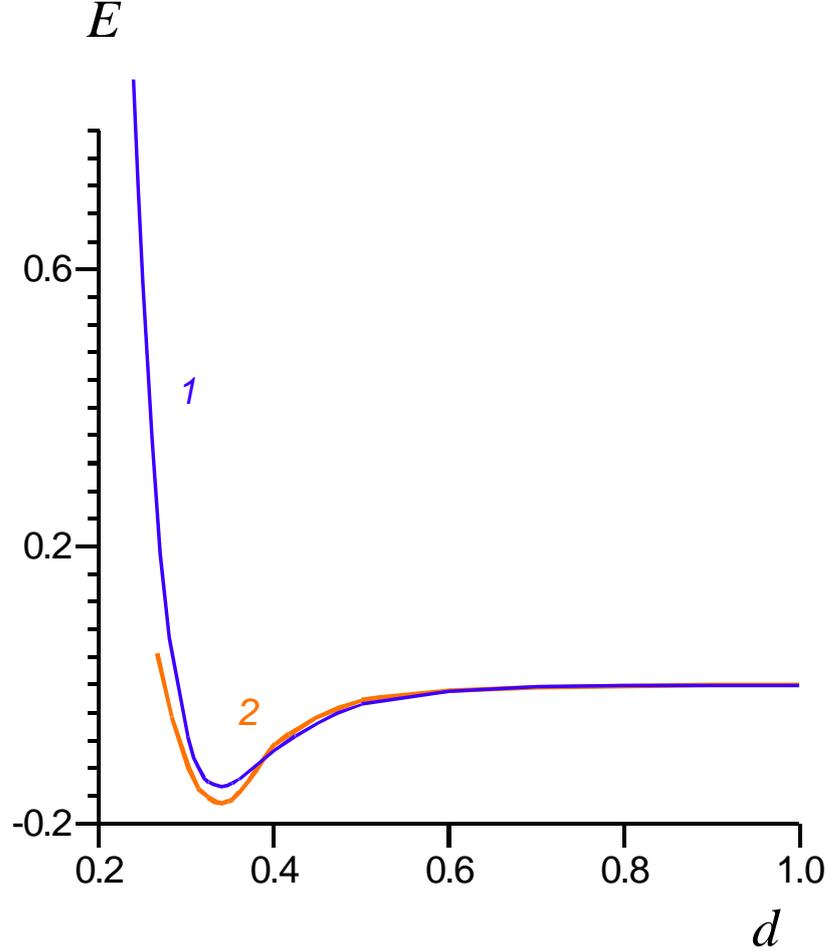

Fig. 4. The total binding energy (eV) per hexagon 1 as a function of the distance $d$ (nm) and 2 – the binding energy per hexagon, calculated by DFT methods taking into account the van der Waals interaction using the Grimme (D3) method

The analysis based on the above methods for graphene is based on the conduction model

$$\sigma(\omega) = \frac{\sigma(0)}{1 + i\omega/\omega_c(\omega)},$$

$$\sigma_{intra}(0,T) = \frac{4\sigma_0 k_B T}{\pi\hbar\omega_{c0}} \ln\left(2 + 2\cosh\left(\frac{\mu_c}{k_B T}\right)\right),$$

where $\sigma_0 = e^2/(4\hbar) = 6.085 \cdot 10^{-5}$ (in S), $\sigma(0) = 4\sigma_0\mu_c/(\pi\hbar\omega_{c0})$, $T$ is the temperature, $\mu_c$ is the chemical potential, $\omega_{c0}$ is the low–frequency CF. We consider the CF to be frequency-dependent, and at zero temperature we have a model

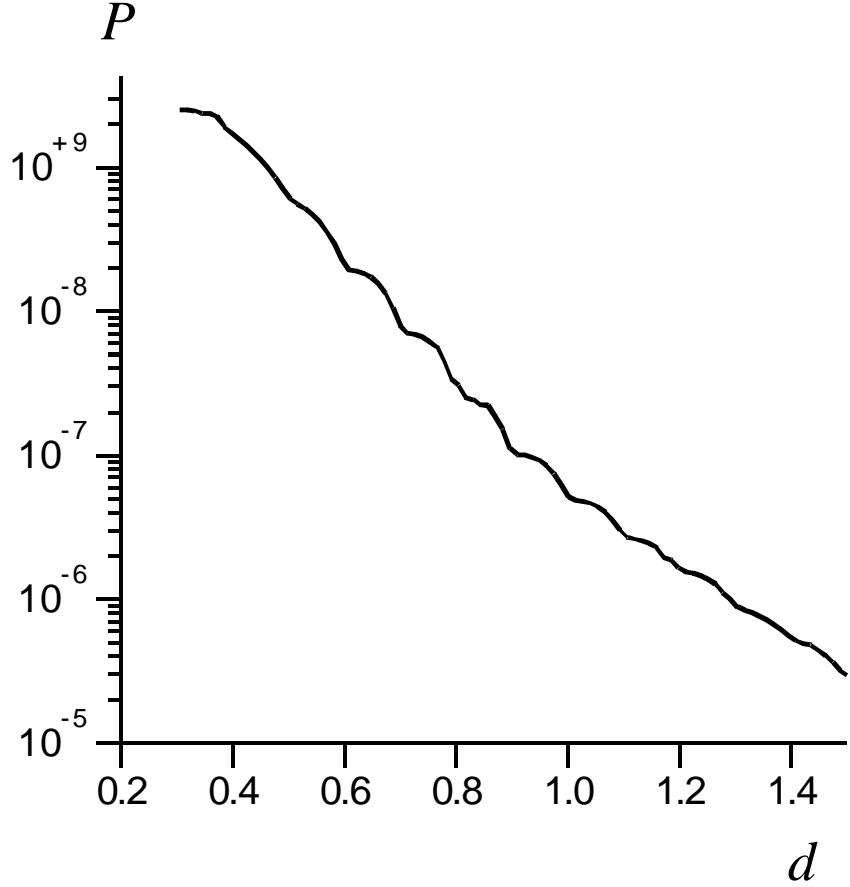

Fig. 5. Van der Waals force density (N/m$^2$) as a function of distance (nm)

$$\omega_c(\omega) = \frac{\omega_{c0}\omega_p^\nu}{\omega_p^\nu + \omega^\nu}.$$

$$\sigma(\omega) = \frac{\sigma(0)\omega_{c0}\omega_p^\nu}{\omega_{c0}\omega_p^\nu + i\omega(\omega_p^\nu + \omega^\nu)}.$$

$$\sigma(0) = 4\sigma_0\mu_c/(\pi\hbar\omega_{c0}).$$

We use $\nu=4$, which ensures convergence. The results are practically independent of it. Figures 4 and 5 show the results for graphene, obtained by the density functional theory method. Calculations using the Van Kampen method are shown in Fig. 6 (see [51]). They correspond well to the calculation of force density at distances of the order of 1 nm or more and show that force is limited at short distances.

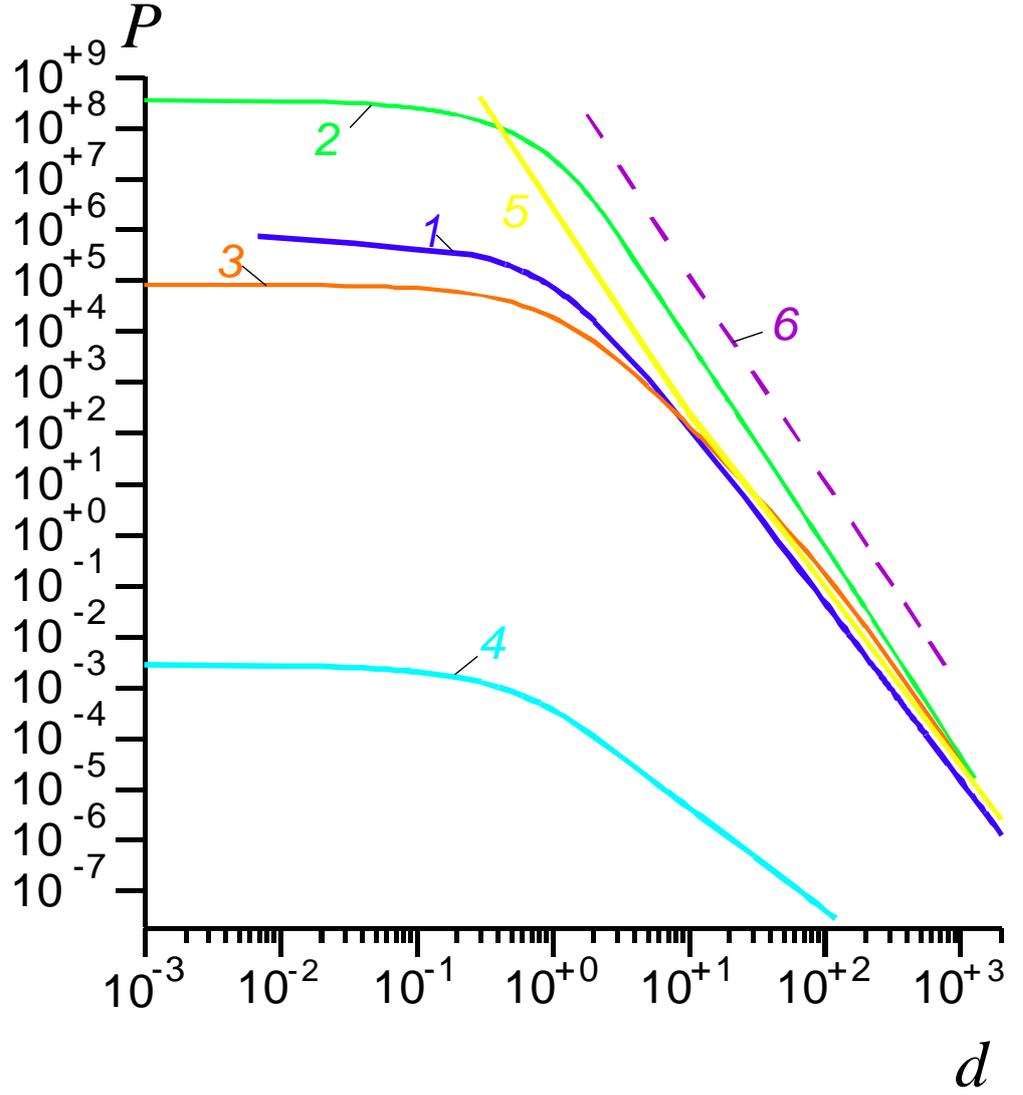

Fig. 6. The density of the Casimir force $F/L^2$ (N/m$^2$) between two graphene sheets as a function of the distance $d$ (nm) according to Van Kampen (7) with $\mu_c = 0.0783$ eV (curve 1), according to the model with DE without a correction factor ($\mu_c = 0.0783$ eV, curve 2) and with correction factors at $\mu_c = 0.1$ (3), $\mu_c = 0.001$ (4). Line 5 is the result from [52]. Dashed line 6 is the result of [13]. CF $\omega_{c0} = 10^{12}$ Hz is everywhere

The Van Kamen method makes it possible to analyze graphene sheets on multilayer substrates. We present the results for two weighted graphene sheets based on the summation of the resonant frequencies in (3). The configuration corresponds to Fig. 1 with the replacement of plates with graphene sheets. The characteristic equation for two sheets in a finite-size resonator has the form

$$y_{mn}^{e,h} + i\left(y_{mn}^{e,h}\left[i\tan(k_z d/2)\right]^{-s} + \varsigma\right)\tan(k_z(D-d)/2) = 0. \tag{25}$$

It contains the normalized (dimensionless) conductivity of graphene $\varsigma = \sigma\eta_0$, $\eta_0 = \sqrt{\mu_0/\varepsilon_0} = 376.78$ (in Ohms) is the characteristic vacuum resistance, $y_{mn}^e = k_0/k_z = k_0/\sqrt{k_0^2 - k_{xm}^2 - k_{yn}^2}$, $y_{mn}^h = k_z/k_0 = \sqrt{k_0^2 - k_{xm}^2 - k_{yn}^2}/k_0$ are the normalized conductivity modes, $s = (-1)^\nu$, $\nu=1$ ($s=-1$) corresponds to the magnetic wall, $\nu=2$ ($s=1$) corresponds to the electric wall. It was solved iteratively for modes E and H, each with electric and magnetic walls (4 modes in total) by finding the quantities $\tilde{k}_{zle} = k_{zle} + \Delta\tilde{k}_{zle}^{e,h}$ and $\tilde{k}_{zlh} = k_{zlh} + \Delta\tilde{k}_{zlh}^{e,h}$, modified by graphene, where $k_{zle} = 2l\pi/D$, $k_{zlh} = (2l-1)\pi/D$. The solutions were taken as

$$1\,\tilde{k}_{zle}^{(e,h)} = \frac{2l\pi}{D} + \Delta\tilde{k}_{zle}^{(e,h)} = \frac{2l\pi}{D-d} + \frac{2\arctan\left(\alpha_{e,h}\left(\tilde{k}_{zle}^{(e,h)},d\right)\right)}{D-d}. \tag{26}$$

$$\tilde{k}_{zlh}^{(e,h)} = \frac{(2l-1)\pi}{D} + \Delta\tilde{k}_{zlh}^{(e,h)} = \frac{(2l-1)\pi}{D-d} + \frac{2\arctan\left(\beta_{e,h}\left(\tilde{k}_{zlh}^{(e,h)},d\right)\right) + \pi}{D-d}. \tag{27}$$

The sum (3) was calculated by limiting the size of the resonator to infinity and was reduced to a two-dimensional integral with respect to $\kappa$ and $k_z=k$. The first integral corresponding to the transverse indices $k_x$ and $k_y$ is transformed in the polar system, where the angle integral is taken elementary. The pressure is determined by differentiating the energy by $d$. The transition to integrals means the appearance of $L_2D$ in the numerator. Therefore, in the iterative solution, we neglect all terms containing powers of $D$ above one in the denominator. The result is

$$P^{\alpha\beta}(d,\delta) = \frac{F^{\alpha\beta}(d,\delta)}{L^2} = \frac{\hbar c}{2\pi^2}\operatorname{Re}\int_0^\infty\int_0^\infty \Phi^{\alpha\beta}(k,d,\delta)\frac{\kappa d\kappa k dk}{\sqrt{\kappa^2+k^2}}, \tag{28}$$

$$\Phi^{\alpha e}(k,d,\delta) = 2\frac{\partial_d \alpha_\alpha(k,d,\delta)}{1+\alpha_\alpha^2(k,d,\delta)} + k,$$

$$\Phi^{\alpha h}(k,d,\delta) = 2\frac{\partial_d \beta_\alpha(k,d,\delta)}{1+\beta_\alpha^2(k,d,\delta)} + k.$$

Here $\alpha,\beta=e,h$, the first indices correspond to the mode, and the second to the wall type. Also

$$\alpha_{e,h}(k,d,\delta) = -\delta\frac{k_0 y_{mn}^{e,h}\tan(kd/2)}{\delta k_0 y_{mn}^{e,h} + \tan(kd/2)}, \tag{29}$$

$$\beta_{e,h}(k,d,\delta) = -\frac{\delta k_0 y_{mn}^{e,h}}{1-\delta k_0 y_{mn}^{e,h}\tan(k_{zlh}d/2)}. \tag{30}$$

The value $\delta=k_0/\varsigma$ is the inverse of conductivity and is related to it by the

$$\delta(\kappa,k) = \frac{k_{c0}k_p^\nu + ik_0\left(k_p^\nu + k_0^\nu\right)}{ik_0 k_{c0}k_p^\nu\xi(0)}, \tag{31}$$

in which wave numbers (plasma and collision) are introduced. Relations (28), (30), obtained similarly to the summation in [13], do not take into account the contribution of evanescent (attenuating) inhomogeneous plane waves, since the value $k_z$ is always real (ignoring dissipation). There are no such waves in the final resonator. They are also not present in the Fabry-Perrault resonator made of perfectly conductive screens. However, in a free space with two graphene sheets, they must be taken into account. They are strongly attenuated in the $z$ direction, so formula (28) should give correct results at large distances. It can be shown that in this case the force is proportional to $1/d^4$. To account for the evanescent contributions, we make a substitution $k = \sqrt{k_0^2 - \kappa^2}$, $kdk = k_0 dk_0$ and transform the integrals to the form

$$\int_0^\infty dk_0 \left( \int_0^{k_0} \Phi^{\alpha\beta}(k_0, \kappa, d, \delta) \kappa d\kappa + \int_{k_0}^\infty \Phi^{\alpha\beta}(k_0, \kappa, d, \delta) \kappa d\kappa \right), \qquad (32)$$

where is in the first integral $k = \sqrt{k_0^2 - \kappa^2}$, and in the second integral $k = -i\sqrt{\kappa^2 - k_0^2} = -iK$. Obviously, the first integral in (32) coincides with (28), and the second gives an addition. Then the modified formulas allow us to calculate the force at short distances. Fig. 7 shows the indicated force without taking into account the evanescent modes, and Fig. 8 shows the calculation for the radiated (curves 1,3,4) and evanescent (2) modes. The sum of both contributions fully corresponds to the results of the Van Kmapen method.

**Conclusion**

There is a very extensive literature on the Casimir effect and the Lifshitz formula, which is difficult to cover, including monographs and reviews, for example, [10,29,53–62]. In most of them, it seems that these are two different approaches. In the work [3] itself, the formula is only given, but it is derived in a large number of publications by different methods. Formally, it is determined from the change in field energy under the action of bodies, i.e. exactly as in the simple Casimir approach for zero oscillations. This raises the problem of complex frequencies in dissipative structures. It is widely discussed in the literature, but little attention has been paid to the method of work [24]. Meanwhile, it allows you to get the right results in dissipative structures, and using an arbitrary number of layers. This is because the principle of the argument gives a valid sum of complex conjugate frequencies.

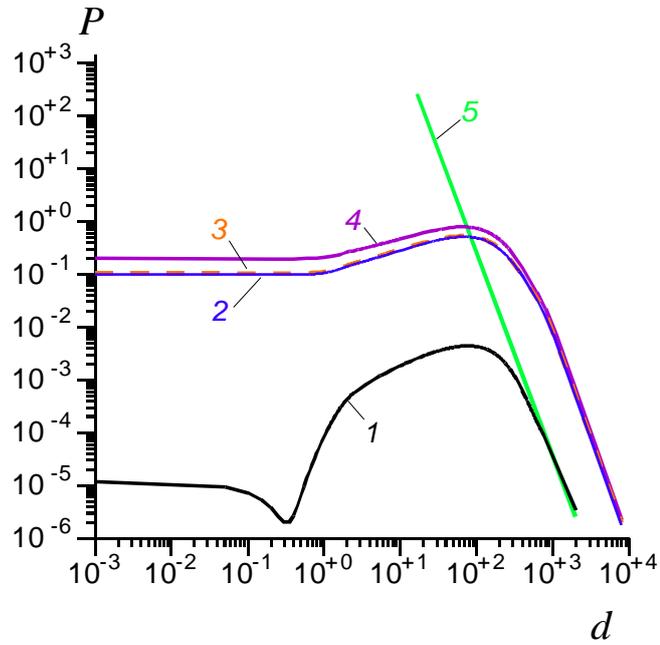

Fig. 7. The density of the Casimir attractive force $P=F/L^2$ (N/m$^2$) between two graphene sheets according to model (28), depending on the distance $d$ (nm) at $\mu_c = 0.0783$ eV (curve 1) and $\mu_c = 7.8$ eV (curves 2–4) and different temperatures: $T=0$, (curve 2), $T=300$ K (3), $T=900$ K (4). Line 5 is the result from [52] at $T=0$. The curves are plotted for $\nu = 4$, $\omega_c = 10^{12}$, Hz

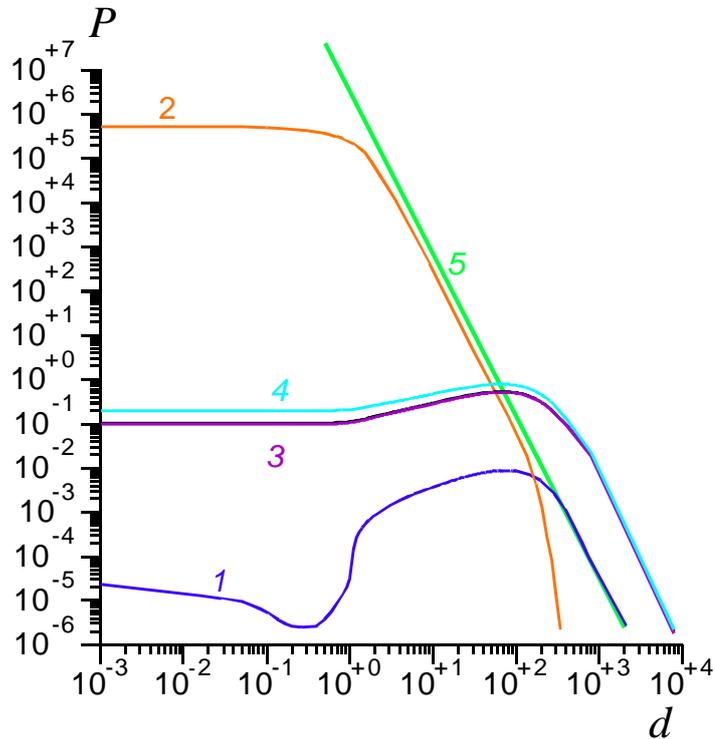

Fig. 8. Densities (N/m$^2$) of the attractive force $\tilde{P}$ (curves 1,3,4) and $P_{ev}$ (2) between two graphene sheets as a function of the distance $d$ (nm) at $\mu_c = 0.0783$ eV (curves 1,2) and $\mu_c = 7.8$ eV (curves 3,4). Curves 1–3 are plotted at a temperature of $T=600$, curve 4 is at $T=900$ (K). Line 5 is the result from [52] at $T=0$. Everywhere $\nu = 4$, $\omega_c = 10^{12}$ Hz

In reality, the DE contains both frequencies. If $\tilde{\omega}_n = \tilde{\omega}'_n + i\tilde{\omega}''_n$ are responsible for absorption (time-damping waves) by atoms, then $\tilde{\omega}_n^* = \tilde{\omega}'_n - i\tilde{\omega}''_n$ correspond to radiation (increasing waves). According to Kirchhoff's law, at thermodynamic equilibrium, radiation at each frequency is exactly equal to absorption, so $\hbar \text{Re}(\tilde{\omega}_n)/2$ is exactly equal to the stored energy, just as $\varepsilon_0 \text{Re}(\varepsilon|\mathbf{E}(\omega)|^2)/2$ is an average stored field energy over the period [4–9] in media where there is no energy accumulation due to particle motion (in this case, always $\text{Re}(\varepsilon) > 1$, unlike plasma, where there may be $\text{Re}(\varepsilon) < 0$ or for a Lorentz oscillator in a narrow range near resonance). In [63], a rather complex approach was proposed for the quantum interaction of a damped oscillator with the thermal field of a thermostat using the Zwanzig–Caldeira–Leggett quantum model. The Van Kampen method makes it possible to circumvent the problem of absorption when calculating Casimir forces. Note that in the Lifshitz formula itself, the real part is taken, but when moving to the plane of the imaginary frequency, the result becomes real. This method is based on DE for PP, which play the main role in dispersion force [64].

**Financing the work**

The work was carried out with the financial support of the Ministry of Education and Science of the Russian Federation within the framework of the state assignment (FSRR-2023-0008).

———————

1. Е.М. Лифшиц, Теория молекулярных сил притяжения между конденсированными телами, ДАН СССР **97**, 4, 643-646 (1954). [E.M. Lifshits, Theory of molecular forces of attraction between condensed bodies, DAN USSR **97**, 4, 643-646 (1954). In Rassian].

2. Е.М. Лифшиц, Влияние температуры на молекулярные силы притяжения между конденсированными телами, ДАН СССР **100**, 5, 879-881 (1955). [E.M. Lifshits, The effect of temperature on the molecular forces of attraction between condensed bodies, DAN USSR **100**, 5, 879-881 (1955). In Rassian].

3. Е.М. Лифшиц, Теория молекулярных сил притяжения между твердыми телами, ЖЭТФ **29**, 94–110 (1955). [E.M. Lifshitz, The Theory of Molecular Attractive Forces between Solids, Sov. Phys. JETP, Vol. 2, 1956, pp. 73–83. Sov. Phys. JETP **2**, 73 (1956)].

4. С.М. Рытов. *Теория электрических флуктуации и теплового излучения* (Изд-во АН СССР, Москва, 1953) [S.M. Rytov. Theory of electrical fluctuations and thermal radiation (Publishing House of the USSR Academy of Sciences, Moscow, 1953). In Russian].

5. М.В. Левин, С.М. Рытов. *Теория равновесных тепловых флюктуаций в электродинамике* (Наука, Москва, 1967). [M.V. Levin, S.M. Rytov. Theory of equilibrium thermal fluctuations in electrodynamics (Nauka, Moscow, 1967). In Russian].

6. С.М. Рытов. *Введение в статистическую радиофизику. Часть 1. Случайные процессы*. (Наука, Москва, 1976). [S.M. Rytov. Introduction to statistical radiophysics. Part 1. Random processes. (Nauka, Moscow, 1976). In Russian]

7. С.М. Рытов, Ю.А. Кравцов, В.И. Татарский. *Введение в статистическую радиофизику. Часть 2. Случайные поля* (Наука, Москва, 1978). [S.M. Rytov, Yu.A. Kravtsov, V.I. Tatarsky. Introduction to statistical radiophysics. Part 2. Random Fields (Nauka, Moscow, 1978). In Russian]

8. И.С. Рыжик, И.М. Градштейн. *Таблицы интегралов, рядов и произведений* (ГИФМЛ (Физматгиз), Москва, 1962). [I.S. Ryzhik, I.M. Gradstein. Tables of integrals, series, and products. (GIFML (Fizmatgiz), Moscow, 1962). In Russian].

9. Л.Г. Прудников, Ю.А. Брычков, О.И. Маричев. *Интегралы и ряды. Элементарные функции* (Наука, Москва, 1981). [L.G. Prudnikov, Yu.A. Brychkov, O.I. Marichev. Integrals and series. Elementary Functions (Nauka, Moscow, 1981). In Russian].

10. W. Simpson, U. Leonhardt, *Forces of the quantum vacuum: an introduction to Casimir physics* (The Weizmann Institute of Science, Israel, 1965).

11. M.V. Davidovich, On energy and momentum conservation laws for an electromagnetic field in a medium or at diffraction on a conducting plate, Phys. Usp. **53** 595–609 (2010). DOI: 10.3367/UFNe.0180.201006e.0623.

12. M.V. Davidovich, On the negative pressure of light in a dispersing medium, Optics and spectroscopy, **131**(9), 1224–1235 (2023). DOI: 10.61011/OS.2023.09.56609.3990-23.

13. H.B.G. Casimir, On the attraction between two perfectly conducting plates, Proc. K. Ned. Akad. Wet. **51**, 793–795 (1948).

14. M.V. Davidovich, O.E. Glukhova. Correlation relations for graphene and its thermal radiation, Izvestiya of Saratov University. Physics, **23**(2), 167–178 (2023). https://doi.org/10.18500/1817-3020-2023-23-2-167-178. [In Russian].

15. J. Schwinger, Casimir effect in source theory, Letters in Mathematical Physics **1**, 43-47 (D. Reidel Publishing Company, 1975).

16. J. Schwinger, L.L. DeRaad, K.A. Milton, Casimir effect in dielectrics, Ann. Phys. **115**(1), 676–698 (1978). DOI: 10.1016/0003-4916(78)90172-0.

17. J. Schwinger, Casimir Effect in Source Theory II, Letters in Mathematical Physics **24**, 59–61 (Kluwer Academic Publishers, 1992).
18. K.A. Milton, Julian Schwinger and the Casimir Effect: The Reality of Zero-Point Energy, arXiv:hep-th/9811054 (1998).
19. F. Intravaia, R. Behunin, Casimir effect as a sum over modes in dissipative systems, Phys. Rev. A **86**, 062517 (2012) DOI: 10.1103/PhysRevA.86.062517.
20. F. Intravaia, How modes shape Casimir physics, International Journal of Modern Physics A **37**(19), 2241014 (2022). DOI: 10.1142/S0217751X22410147.
21. I.E. Dzyaloshinskii, E.M, Lifshitz, L.P. Pitaevskii, General theory of van der Waals's forces, Sov. Phys. Usp. **4** 153–176 (1961); DOI: 10.1070/PU1961v004n02ABEH003330.
22. I.E. Dzyaloshinskii, E.M, Lifshitz, L.P. Pitaevskii, The general theory of van der Waals forces, Advances in Physics, **10** (38). 165–209 (1961). DOI: 10.1080/00018736100101281.
23. I.E. Dzyaloshinskii, E.M. Lifshitz, L.P. Pitaevskii, Van Der Waals Forces in Liquid Films, Soviet Phys. JETP **10**, 161–170 (1960).
24. N.G. Van Kampen, B.R.A. Nijboer, K. Schram, On the macroscopic theory of van der Waals forces, Phys. Lett. A **26**, 307–308 (1968). DOI: 10.1016/0375-9601(68)90665-8.
25. K. Schram, On the macroscopic theory of retarded Van der Waals forces, Physics Letters A **43**(3), 282–284 (1973). DOI: 10.1016/0375-9601(73)90307-1%20.
26. B.W. Ninham, V.A. Parsegian, G.H. Weiss, On the macroscopic theory of temperature-dependent van der Waals forces J. Statistical Physics 2(4):323-328 DOI: 10.1007/BF01020441.
27. Volokitin A.I., Persson B.N.J. Electromagnetic Fluctuations at the Nanoscale. Theory and Applications. (Heidelberg: Springer, 2017).
28. A.I. Volokitin, B.N.J. Persson, Radiative heat transfer and noncontact friction between nanostructures, Phys. Usp. **50** 879–906 (2007). DOI: 10.1070/PU2007v050n09ABEH006192.
29. S.K. Lamoreaux, The Casimir force: Background, experiments and applications, Reps. Progr. Phys. **65**, 201–236, (2005). DOI:10.1088/0034-4885/68/1/R04.
30. P.W. Milonni, *The Quantum Vacuum* (San Diego, CA: Academic, 1994).
31. А.Л. Фельдштейн, Л.Р. Явич. *Синтез четырехполюсников и восьмиполюсников на СВЧ* (Связь, Москва, 1971). [A.L. Feldstein, L.R. Yavich. Synthesis of four-pole and eight-pole microwave devices (Svyaz, Moscow, 1971). In Russian],

32. K. Schram, G.L. Klimchitskaya, U. Mohideen, V.M. Mostepanenko, The Casimir force between real materials: Experiment and theory, Rev. Mod. Phys. 81, 1827 (2009). DOI: 10.1103/RevModPhys.81.1827.
33. F.S.S. Rosa, D.A.R. Dalvit, P.W. Milonni, Electromagnetic energy, absorption, and Casimir forces: Uniform dielectric media in thermal equilibrium, Phys. Rev. A **81**, 033812, 2010 DOI: 10.1103/PhysRevA.81.033812.
34. F.C. Lombardo, F.D. Mazzitelli, A.E. Rubio López, Casimir force for absorbing media in an open quantum system framework: Scalar model, Phys. Rev. A 84, 052517, 2011 DOI: 10.1103/PhysRevA.84.052517.
35. F.S.S. Rosa, D.A.R. Dalvit1, P.W. Milonni, Electromagnetic energy, absorption, and Casimir forces. II. Inhomogeneous dielectric media, Phys. Rev. A 84, 053813, 2011 DOI: 10.1103/PhysRevA.84.053813.
36. T.G. Philbin, Canonical quantization of macroscopic electromagnetism, *New J. Phys.* **12**, 123008 (2010). DOI: 10.1088/1367-2630/12/12/123008.
37. T.G. Philbin Casimir effect from macroscopic quantum electrodynamics, New J. Phys. **13** 063026 (2011). DOI:10.1088/1367-2630/13/6/063026.
38. R.O. Behunin, B.-L. Hu, Nonequilibrium forces between atoms and dielectrics mediated by a quantum, field Phys. Rev. A **84**, 012902 (2011). DOI: 10.1103/PhysRevA.84.012902.
39. А.И. Ахиезер И.А. Ахиезер, *Электромагнетизм и электромагнитные волны* (Высшая школа, Москва, 1985). [A.I. Akhiezer I.A. Akhiezer, Electromagnetism and electromagnetic waves (Higher School, Moscow, 1985). In Russian].
40. Л.Д. Гольдштейн, Н.В. Зернов. *Электромагнитные поля и волны* (Советское радио, М., 1971). [L.D. Goldstein, N.V. Zernov. Electromagnetic fields and waves (Soviet Radio, Moscow, 1971). In Russian].
41. Л.А. Вайнштейн, *Электромагнитные волны* (Радио и связь, Москва, 1988). [L.A. Wainstein, Electromagnetic waves (Radio and Communications, Moscow, 1988). In Russian].
42. M.V. Davidovich, Electromagnetic Energy Density and Velocity in a Medium with Anomalous Positive Dispersion, Technical Physics Letters, **32**(11), 982–986 (2006). © Pleiades Publishing, Inc., DOI: 10.1134/S106378500611023X.
43. M.V. Davidovich, On the Electromagnetic Energy Density and Energy Transfer Rate in a Medium with Dispersion due to Conduction, Technical Physics **55**(5), 630–635 (2010). © Pleiades Publishing, Ltd.. DOI: 10.1134/S1063784210050063.

44. M.V. Davidovich, On Times and Speeds of Time-Dependent Quantum and Electromagnetic Tunneling, Journal of Experimental and Theoretical Physics **130**(1), 35–51 (2020). © Pleiades Publishing, Inc.. DOI: 10.1134/S1063776119120161.

45. Bryksin, M.P. Petrov, Casimir force with the inclusion of a finite thickness of interacting plates, Phys. Solid State **50**(2), 229–234 (2008). DIO: 10.1134/S1063783408020029.

46. M.V. Davidovich, Plasmons in multilayered plane-stratified structures, Quantum Electronics **47**(6), 567–579 (2017). DOI: 10.1070/QEL16272.

47. S.V. Gaponenko, D.V. Novitsky, Wigner time for electromagnetic radiation in plasma, Phys. Rev. A 106, 023502 (2022). DOI: 10.1103/PhysRevA.106.023502.

48. V.V. Sobolev, A.P. Timonov, V.Val. Sobolev, Fine Structure of the Dielectric-Function Spectrum in Diamond, Semiconductors, **34**(8), 2000, 902–907 (2000). Translated from Fizika i Tekhnika Poluprovodnikov, Vol. 34, No. 8, 2000, pp. 940–946. DOI: 10.1134/1.1188098.

49. T.L. Cocker, D. Baillie, M. Buruma, L.V. Titova, R.D. Sydora, F. Marsiglio, F.A. Hegmann, Microscopic origin of the Drude-Smith model, Phys. Rev. B **96**, 205439 (2017). DOI: 10.1103/PhysRevB.96.205439.

50. W.-C. Chen, R.A. Marcus, The Drude-Smith Equation and Related Equations for the Frequency-Dependent Electrical Conductivity of Materials: Insight from a Memory Function Formalism, Chemphyschem **22**(16),1667–1674 (2021). DOI: 10.1002/cphc.202100299.

51. M.V. Davidovich. Dispersion interaction of two graphene sheets, arXiv:2508.17999. https://doi.org/10.48550/arXiv.2508.17999 (2025).

52. A.I. Volokitin, B.N.J. Persson, Effect of the Electric Current on the Casimir Force between Graphene Sheets, JETP Letters, **98**(3), 143–149 (2013). © Pleiades Publishing, Inc. DOI: 10.1134/S0021364013160145.

53. Yu.S. Barash, V.L. Ginzburg, Electromagnetic fluctuations in matter and molecular (Van-der- Waals) forces between them, Sov. Phys. Usp. **18,** 305–322 (1975). DOI: 10.3367/UFNr.0116.197505a.0005.

54. K.A. Milton, *The Casimir Effect: Physical Manifestations of Zero-Point Energy* (Singapore: World Scientific, 2001).

55. G.L. Klimchitskaya, A.B. Fedortsov, Y.V. Churkin, V.A. Urova, Casimir force pressure on the insulating layer in metal-insulator-semiconductor structures. Phys. Solid State **53**, 1921–1926 (2011). DOI: 10.1134/S1063783411090174.

56. M. Bordag (ed.), *The Casimir Effect 50 Years Later* (World Scientific, Singapore, 1999).

57. B. Geyer, G.L. Klimchitskaya, V.M. Mostepanenko, Thermal quantum field theory and the Casimir interaction between dielectrics, Phys. Rev. D 72, 085009 (2005). DOI: 10.1103/PhysRevD.72.085009.
58. G. Plunien, B. Muller, W. Greiner, The Casimir effect, Phys. Rep. **134**, 87 (1986). DOI:10.1016/0370-1573(86)90020-7.
59. M. Bordag, U. Mohideen, and V. M. Mostepanenko, New developments in the Casimir effect, Phys. Rep. **353**, 1 (2001). DOI: 10.1016/S0370-1573(01)00015-1.
60. K.A. Milton, The Casimir effect: Recent controversies and progress, J. Phys. A **37**, R209 (2004). **DOI** 10.1088/0305-4470/37/38/R01.
61. M. Bordag, G. Klimchitskaya, U. Mohideen, V. Mostepananko, *Advances in the Casimir effect*, Int. Ser. Monogr. Phys. 145, 1 (2009). https://inspirehep.net/ literature/841474.
62. V.M.Mostepanenko, N.N. Trunov, The Casimir effect and its applications, Sov. Phys. Usp. **31** 965–987 (1988). DOI: 10.1070/PU1988v031n11ABEH005641.
63. Yu.S. Barash. Damped Oscillators within the General Theory of Casimir and van der Waals Forces. Journal of Experimental and Theoretical Physics **132**(4), 663-674 (2021). DOI: 10.1134/S1063776121040014.
64. M.V. Davidovich, Casimir-Lifshitz force and plasmons in a structure with two graphene sheets, Proc. SPIE **11846**, Saratov Fall Meeting 2020: Laser Physics, Photonic Technologies, and Molecular Modeling, 118460J (2021). https://doi.org/10.1117/12.2589911.

**Figures**

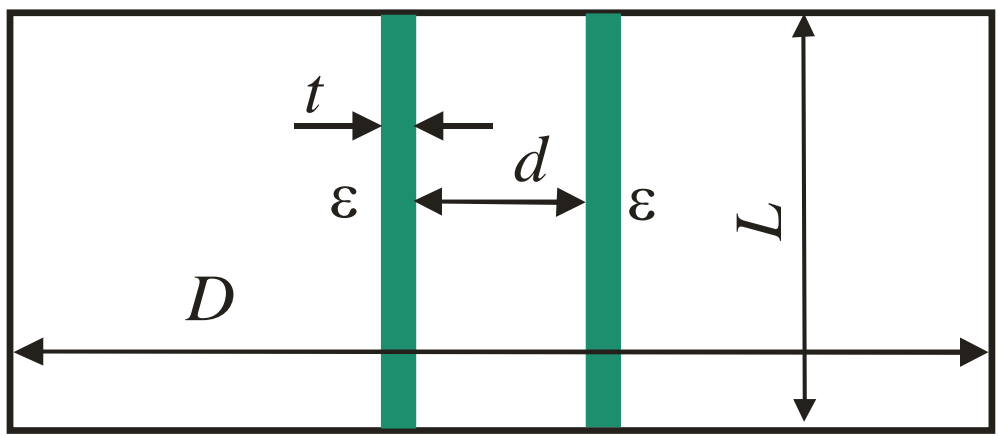

Fig. 1

Rectangular resonator with two dielectric layers with $L_x=L_y=L$ and $L_z=D$

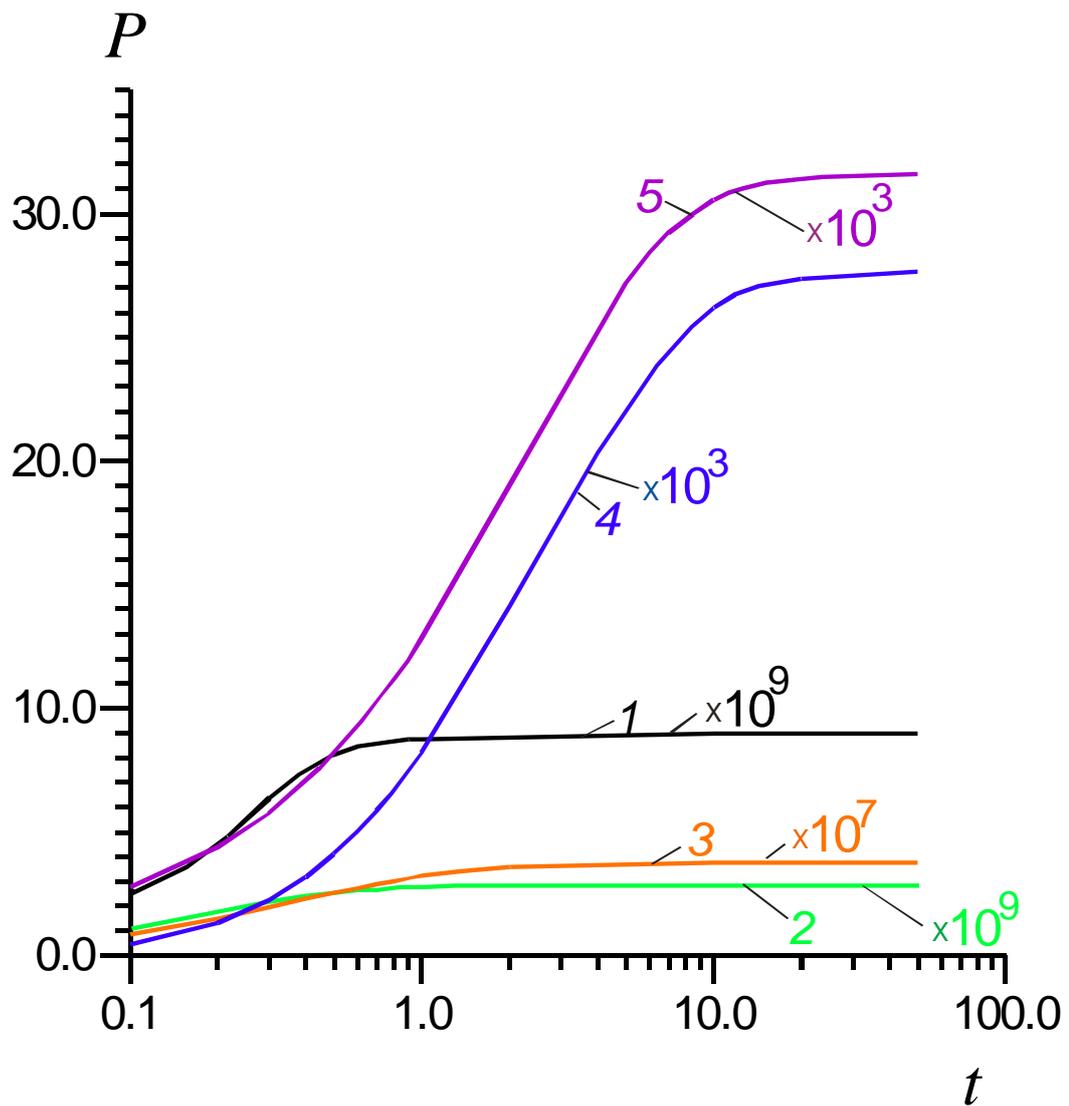

Fig. 2.

Casimir pressure $P=F/L^2$ (N/m$^2$) between two dielectric layers depending on their thickness $t$ (nm) at different distances $d$ (nm): $d=0.01$ (curve 1); 0.1 (2); 1 (3); 10 (4,5). Curves 1-4 are plotted in the absence of conductivity ($\omega_p = 0$), curve 5 – in the presence of it

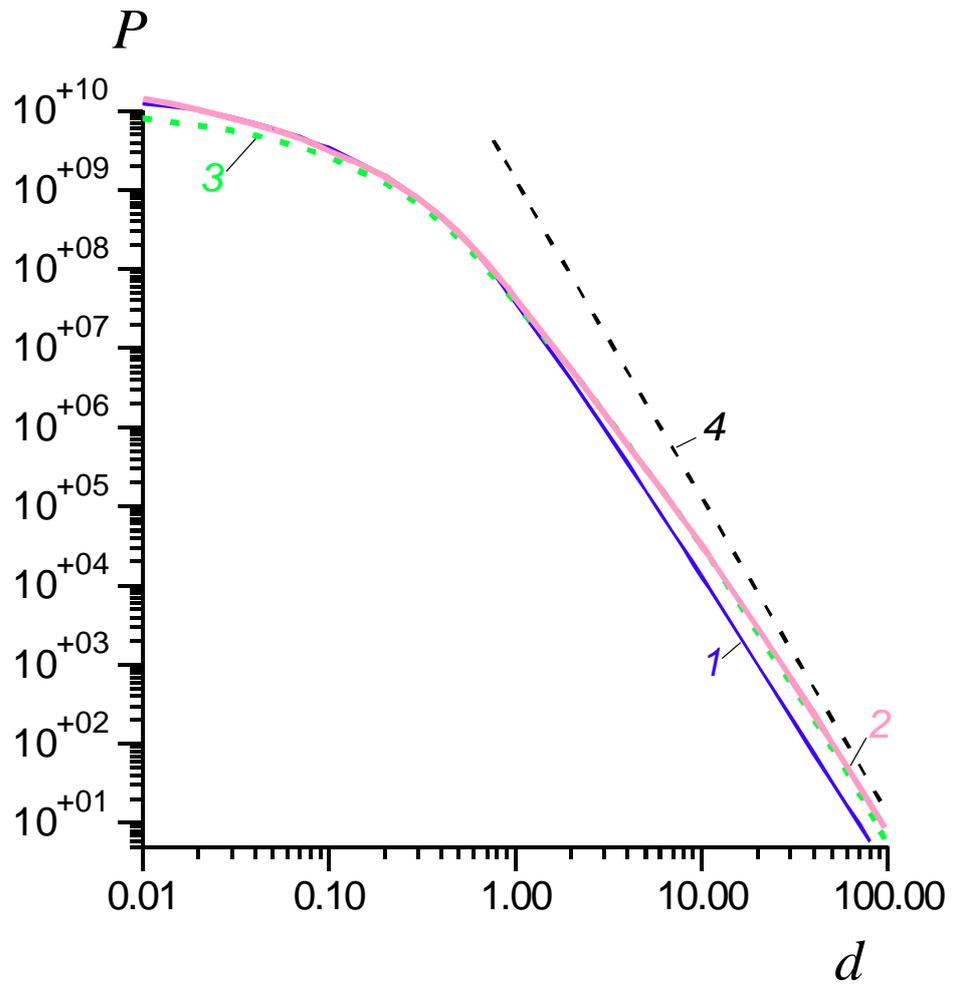

Fig. 3.

Casimir pressure P (N/m$^2$) between two dielectric layers depending on the distance d (nm) at different thicknesses $t$ (nm): $t =1$ (curve 1); 50 (2.3). Curves 1, 2 are constructed taking into account the conductivity, curve 3 at $\omega_p = 0$. 4 – the Casimir result

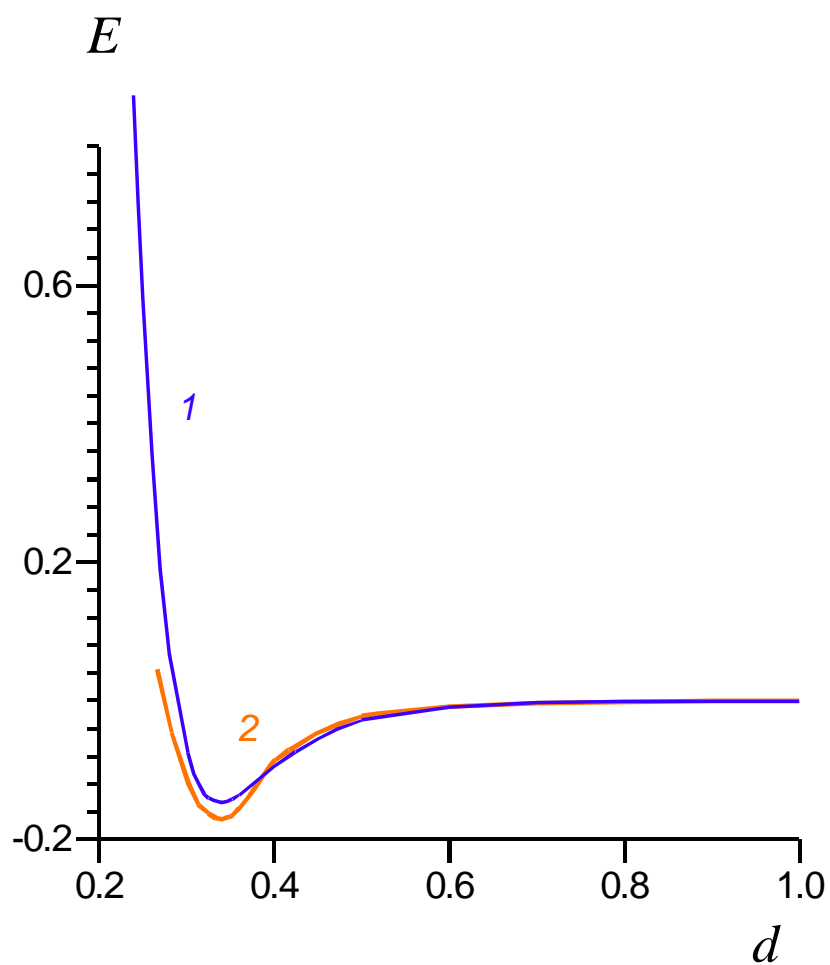

Fig. 4.

The total binding energy (eV) per hexagon 1 as a function of the distance $d$ (nm) and 2 – the binding energy per hexagon, calculated by DFT methods taking into account the van der Waals interaction using the Grimme (D3) method

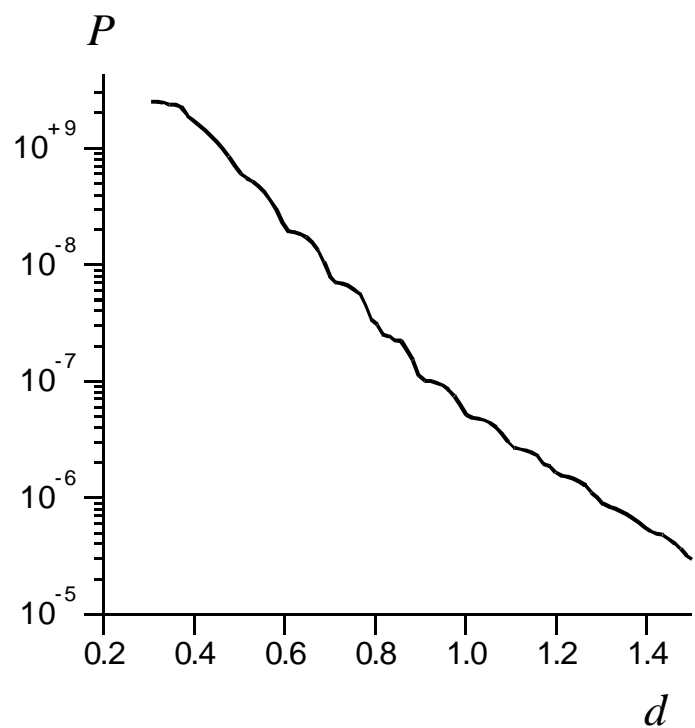

Fig. 5.

Van der Waals force density (N/m$^2$) as a function of distance (nm)

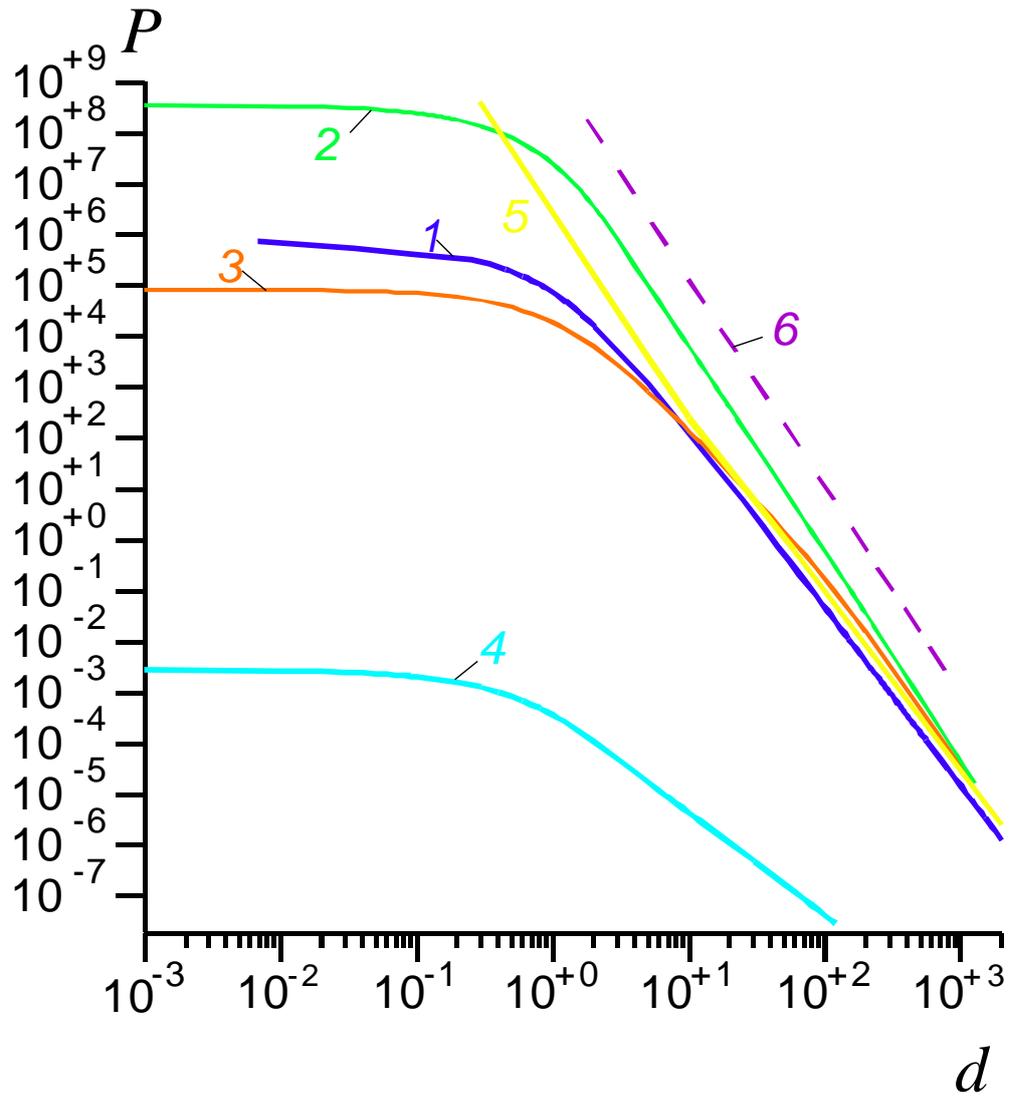

Fig. 6.

The density of the Casimir force $F/L^2$ (N/m$^2$) between two graphene sheets as a function of the distance $d$ (nm) according to Van Kampen (7) with $\mu_c = 0.0783$ eV (curve 1), according to the model with DE without a correction factor ($\mu_c = 0.0783$ eV, curve 2) and with correction factors at $\mu_c = 0.1$ (3), $\mu_c = 0.001$ (4). Line 5 is the result from [51]. Dashed line 6 is the result of [13]. CF $\omega_{c0} = 10^{12}$ Hz is everywhere

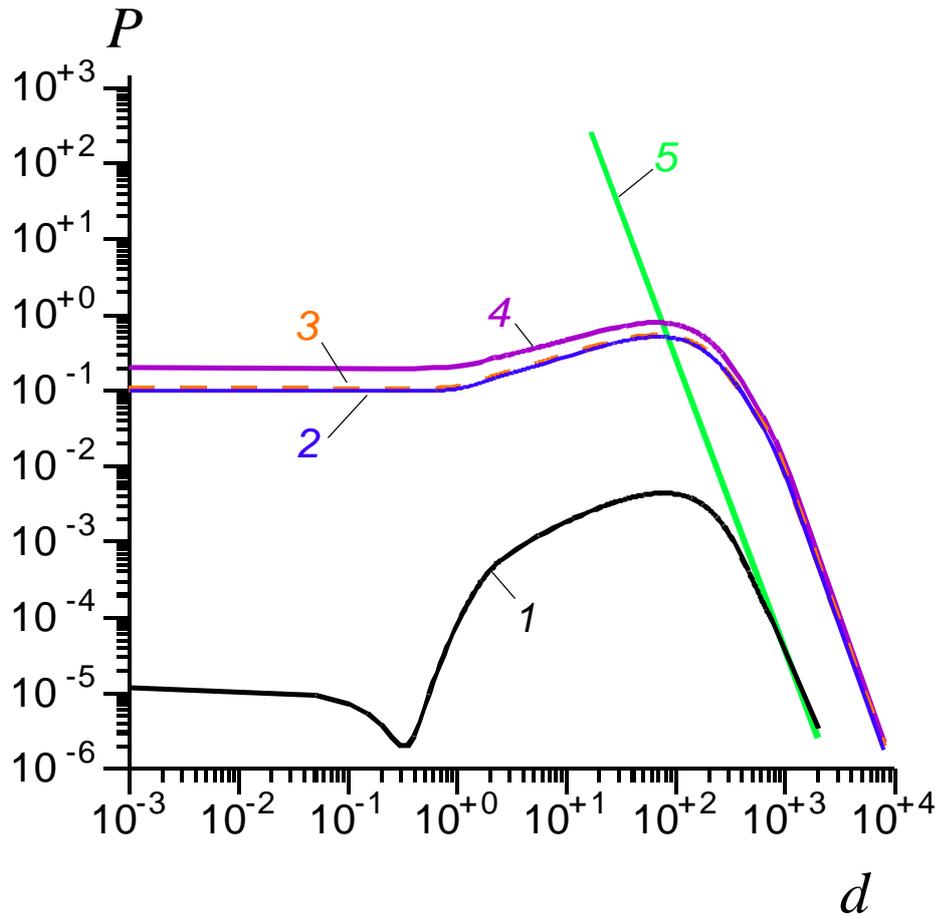

Fig. 7.

The density of the Casimir attractive force $P=F/L^2$ (N/m$^2$) between two graphene sheets according to model (28), depending on the distance $d$ (nm) at $\mu_c = 0.0783$ eV (curve 1) and $\mu_c = 7.8$ eV (curves 2-4) and different temperatures: $T=0$, (curve 2), $T=300$ K (3), $T=900$ K (4). Line 5 is the result from [51] at $T=0$. The curves are plotted for $\nu = 4$, $\omega_c = 10^{12}$, Hz

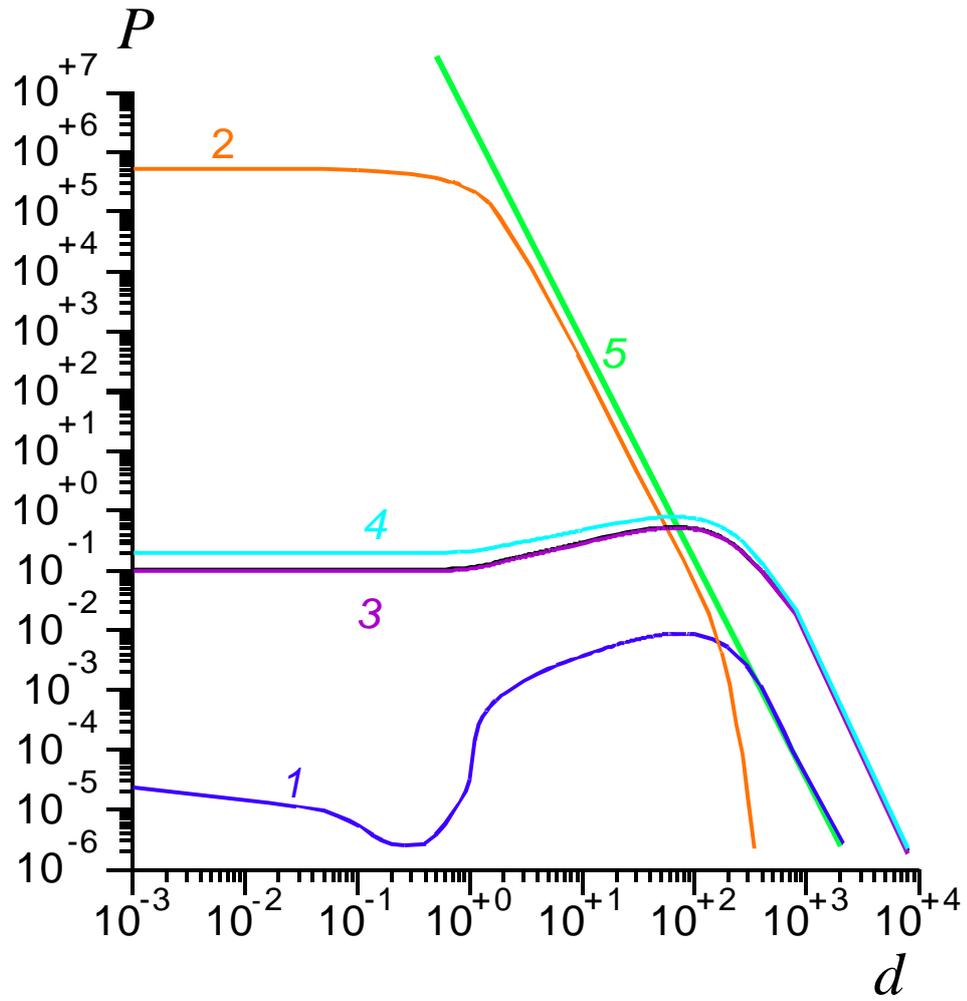

Fig. 8.

Densities (N/m$^2$) of the attractive force $\tilde{P}$ (curves 1,3,4) and $P_{ev}$ (2) between two graphene sheets as a function of the distance $d$ (nm) at $\mu_c = 0.0783$ eV (curves 1,2) and $\mu_c = 7.8$ eV (curves 3,4). Curves 1-3 are plotted at a temperature of $T=600$, curve 4 is at $T=900$ (K). Line 5 is the result from [51] at $T=0$. Everywhere $v=4$, $\omega_c = 10^{12}$ Hz